\documentclass[footinbib]{article}

\usepackage[font={small}, labelfont=bf, justification=centering]{caption}[2007/01/07]
\usepackage[numbers]{natbib}
\usepackage[utf8]{inputenc}
\usepackage{mathtools}
\usepackage{amsmath}
\usepackage{amsbsy}
\usepackage{caption}
\usepackage{breqn}
\usepackage{amssymb}
\usepackage{bm}
\usepackage{units}
\usepackage{cleveref}
\usepackage{color,soul}
\usepackage{float}
\usepackage{tikz}
\usepackage{enumitem}
\usepackage{afterpage}
\usepackage{booktabs}
\usepackage{multirow}
\usepackage{needspace}
\usepackage{xcolor}

\usepackage[LY1]{fontenc}
\usepackage[lf]{berenis}

\DeclareTextCompositeCommand{\k}{LY1}{a}
  {\oalign{a\crcr\noalign{\kern-.27ex}\hidewidth\char7}}

\usepackage{changes}
\definechangesauthor[name={Per cusse}, color=orange]{per}
\setremarkmarkup{(#2)}

\DeclareMathAlphabet\mathbfcal{OMS}{cmsy}{b}{n}

\usepackage{mymacros_10Sep19}
\usepackage{xargs}                      
\usepackage{fullpage,lmodern}
\usepackage{array}
\usepackage{epstopdf}
\usepackage[normalem]{ulem}

\usepackage{fullpage,lmodern}
\usepackage[final]{showlabels,rotating} 

\graphicspath{ {Fig/} }

\crefname{item}{A.}{A.s}

\newcommandx{\rld}[2][1=]{\todo[inline,linecolor=blue,backgroundcolor=red!25,bordercolor=green,#1]{RLD: #2}}

\crefname{equation}{Equation}{Equations}
\crefname{table}{Table}{Tables}
\crefname{figure}{Figure}{Figures}
\crefname{section}{Section}{Sections}

\usepackage{authblk}

\begin{document}

\title{A unified description of Surface Free Energy and Surface Stress}
\author{\footnote{Corresponding author} Nicodemo Di Pasquale }
\author{Ruslan L. Davidchack}
\affil{School of Mathematics and Actuarial Science$,$ University of Leicester$,$ University Rd$,$ Leicester LE1 7RH$,$ UK}
\date{}
%
%

\maketitle

\begin{abstract}
\noindent Even though the study of interfacial phenomena dates back to Laplace and was formalised by Gibbs, it appears that some concepts and relations among them are still causing some confusion and debates in the literature, particularly for interfaces involving solids.  Moreover, ever since the Molecular Dynamics (MD) simulations have started to be widely used in the study of surface properties, these debates only intensified.  In this work, we present a systematic description of the interfacial properties from the thermodynamic and statistical mechanics points of view. In particular, we link our derivations to MD simulations, describing precisely what different quantities represent and how they can be calculated. 
We do not follow the usual way that consists of describing the thermodynamics of the surfaces in general and then considering specific cases (e.g. liquid-liquid interface, liquid-solid interface). Instead, we present our analysis of various properties of surfaces in a hierarchical way, starting with the simplest case that we have identified: a single component liquid-vacuum interface, and then adding more and more complications when we progress to more complex interfaces involving solids. We propose that the term ``surface tension'' should not be used in the description of surfaces and interfaces involving solids, since its meaning is ambiguous.  Only ``Surface Free Energy'' and ``Surface Stress'' are well defined and represent distinct, but related, properties of the interfaces.  We demonstrate that these quantities, as defined in thermodynamics and measured in MD simulations, satisfy the Shuttleworth equation. 
\end{abstract}

\section{Introduction}\label{Sec:primaintro}

\noindent The surface of any object is usually viewed as the limiting boundary that separates two different regions of space: its inside and its outside. We call the inside of the object \textit{bulk} and we only consider the case when the composition of the bulk is homogeneous. The external region can be vacuum or another bulk phase in contact with the first one. In the latter case, we will also consider this second phase to be homogeneous.  Within this picture, the external surface represents the interface between these regions, and the great attention that is devoted to the study of its properties comes from the fact that it behaves very differently from the bulk region of the volume of space that it encloses. In fact, a wide range of phenomena (i.e. adsorption \citep{Dkabrowski2001}, adhesion \citep{Kendall1971,Johnson1971,Packham2003}, aggregation \citep{Berrill2011}, capillarity \citep{Washburn1921}) are related to the properties of the interface.

Probably, the most known description of an interface is the one due to Gibbs \citep{Gibbs1957}. Gibbs's analysis of the interface between two different phases is obtained by defining a zero-width \textit{dividing surface} and ascribing all the differences in thermodynamic quantities between the interface and the bulk to the excess quantities characterising this dividing surface \citep{Gibbs1957}. The strength of this description lies in the fact that, once the position of the dividing surface is specified within the system, excess quantities can be uniquely and intuitively defined. However, the main problem with this approach is related to the positioning of the surface and the dependence of the excess quantities on this position. Moreover, this surface is a two-dimensional object, i.e., with zero thickness, which does not represents a real physical system, where the presence of the interface influences the nearby regions, thus giving thickness to the interfacial region.  While Gibbs' mathematical model is logically coherent and mathematically sound, there is no such thing as a zero-width interface in real systems where two phases, $\alpha$ and $\beta$, are in contact. What we observe is a \textit{transition} region with a small, but non-zero, thickness over which material properties change gradually from their values in the bulk of the first phase (e.g. $\alpha$) to their values in the bulk of the second phase (e.g. $\beta$). 

With the introduction of the computers, the numerical integration of the equations of motion was possible for every atom of a small sample of a specific substance, a methodology which became known as Molecular Dynamics (MD) simulations  \cite{Frenkel2001}. Thanks to this, MD simulations give a realistic model of the system under study, allowing the calculation of any property of interest from its statistical mechanics definition.
With MD simulations we can model a small portion of material, including its interface. Here, the Gibbs approach does not link naturally with the atomistic description of materials. For this reason, a model that reflects better the real description of the systems with an interfaces, as finite volumes of material with changing properties with respect the bulk, is more appropriate in the context of MD simulations. One of such models was proposed by \citet{Guggenheim1985} and will be considered in the next sections as a starting point for the derivations of the surface properties in different systems. This  approach, in fact, focusing on a small region of the sample results more suitable for the framework of MD simulations. Within MD simulations we can put two phases in contact and simulate the interface layer, obtaining properties directly from the statistical mechanics definitions, as averages of various microscopic quantities over a finite volume of space and over time.

In this work we revise the theory of the interfaces, starting from the thermodynamic point of view and we show how it is linked with statistical mechanics definition of the relevant quantities describing the interfaces. \\ 
Our goal is to present a consistent theory for MD simulations where the surfaces quantities are clearly and rigorously defined. For this reason, different way to calculate interfaces quantities in MD simulations are analysed in the context of the theory we are developing. We also present a new derivation of the Shuttleworth equation \cite{Shuttleworth1950}, based on statistical mechanics considerations.

The paper is structured as follows: we start with the macroscopic approach, by describing the systems using classical thermodynamic concepts using the \citet{Guggenheim1985}  approach of the finite volumes of materials, instead of that of Gibbs of the dividing surface. We then introduce the relevant statistical mechanics quantities and we show the links between the two descriptions. After that, we focus on the MD simulations, and in particular the cleaving model to calculate the Surface Free Energy of a system. We then show some results on a test system represented by a Lennard-Jones crystal and we draw some conclusions.

\section{Description of the problem}\label{Sec:intro}

Before starting with the analysis of the interface we state the following working assumptions:
\begin{enumerate}[label=\roman*]
	\item The phases are at rest;
	\item Flat surface. Without loss of generality, the sample is oriented with the $z$-direction always perpendicular to the exposed surface; \label{hyp:flat}
	\item Complete equilibrium: \label{hyp:eq}
		\begin{enumerate}[label=\alph*]
			\item  \textit{Thermal equilibrium}: temperature $T$ is uniform throughout the system ($\alpha +\beta$);
			\item \textit{Diffusive Equilibrium}: no net flux of materials within the system;
			\item \textit{Mechanical Equilibrium}: if both phases in contact are fluid phases, the pressure $P$ is uniform throughout the system. If a solid phase is present, we consider the solid oriented such that the direction (it will always be the $z$ direction) of one of the principal stresses is aligned with the normal to the interface (for a definition of principal stresses in solids see \cref{Sec:SolInt}); \label{hyp:me}
		\end{enumerate}
	\item No chemical reactions within the system;
	\item All the systems considered are single-component.
\end{enumerate}
In our work we will use two conventions to identify a point in space, $\bfr = (\x, \y, \z)$ and $\bfx=(x,y,z)$, whichever is more convenient in our discussion. It is always assumed that $\x$ is equivalent to the $x$-direction and analogously for $y$ and $z$. 

The diversity of terms and definitions used to describe the thermodynamic state of the material surface have created a lot of misunderstanding in the literature throughout the years.  Before starting our discussion, we want to give clear definitions of the quantities we are going to analyse. The terms {\em surface tension} and {\em surface stress} are sometimes used as synonyms. However, their meaning is quite different with respect the thermodynamic theory and this fact must be always kept in mind. Surface tension usually refers to liquids and represents the reversible work required to create a unit area of surface \citep{Vermaak1968}. This definition suggests that a more appropriate name for this quantity should be related to the fact that it represents energy, being also measured in the units of energy per area (e.g., mJ/m$^2$ in the SI). In particular, it represents a thermodynamic {\em free energy} and we will call it Surface Free Energy (SFE).

The reason why the SFE is sometimes referred to as ``surface tension'' lies in its mechanical interpretation \cite{Brown1947}. Any material system evolves to reduce its Gibbs free energy, and in systems with a surface this results in the minimisation of the total surface area (the molecular reason will be explained in \cref{Sec:MolExpl}). This is the essence of the Wullf theorem and is the reason why a droplet in isotropic conditions assumes a spherical shape, the sphere being the geometrical object with the least surface area for a given volume \cite{Herring1951,Wulff1901}. 
In a system with a flat surface, in order to balance this tendency, a tensile force tangent to the surface must be applied at its edges. This state of tension of the surface, which was fully accepted not so long ago \citep{Brown1947}, can be easily observed \citep{Jasper1972}.  If we consider a line on the surface, surface tension represents this tangent force in the direction normal to the line divided by the length of the line over which the force acts. In this case it is naturally measured in the units of force per length (e.g., N/m in the SI). One important feature that results from this definition, which comes from the circular symmetry of the surface in liquids, is that we have not specified any particular direction of this line with respect to the surface. 
That is, the SFE, or surface tension, as defined in liquids, is a scalar quantity. 
 
The concept of {\em surface stress} (SS), first introduced by Gibbs, is different from the SFE and is related to the fact that the area of a surface can be increased by reversible (elastic) stretching of the pre-existing surface.  The main difference between SS and SFE is that liquids cannot be elastically stretched (see the end of \cref{Sec:MolExpl} for a short discussion on this). Therefore, the concept of SS does not apply to them.  The term surface tension can suggest some similarity with the state of stress within the solid that is the cause of the surface stress, but, as we said above, the term ``surface tension'' can be used unambiguously to describe only liquid interfaces.  In solids, the term ``surface tension'' is used by different authors to mean either the mechanism of elastic stretching or the reversible creation of the interface \citep{Orowan1970,Fletcher2014,Gloor2005,Hui2013}, which often leads to confusion.  For this reason, following the suggestion of \citeauthor{Cahn1979} \citep{Cahn1979}, we believe that the term surface tension should be abandoned in favour of the more precise and uniquely defined term SFE for the reversible work of unit surface area creation, and the term SS for the work per unit area in the case of area creation by elastic stretching of a pre-existing surface. The importance of this distinction between the two concepts lies in the fact that not only crystalline materials (i.e. the ones usually considered in condensed matter) can be elastically stretched, but also a wide class of systems in the soft matter field, e.g. cross-linked polymer network \citep{Andreotti2016,Andreotti2016b} or polymer gels \cite{Xu2018}.

In the following we will need to describe the volume and area variation of a sample and we introduce here the notations to describe its elastic deformation. 
Let's denote by $\bfr = (\x, \y, \z)$ the position of a material point in an unstrained solid, and by $\hbfr = (\hat{x}{}^1, \hat{x}{}^2,\hat{x}{}^3)$ its position in the solid after deformation. The difference $\e_i = \hat{x}{}^i - x^i$, $i = 1,2,3$, measures the displacement of the point due to the deformation. Note that the coordinates $\hbfr$ of the displaced point are functions of the coordinates $\bfr$ of the point before the deformation.  For small deformation, a strain tensor $\strain$, with components $\e_{ij}$, is defined as \citep{Landau1959}:
\begin{equation}
	\e_{ij} = \frac{1}{2}\cip{\pard{\e_i}{x^j}+\pard{\e_j}{x^i}}, \;\; i,j=1,2,3.
\end{equation} 
In this work, we indicate the infinitesimal strain as $\e_{ij}$ instead of $\delta \e_{ij}$ to avoid cumbersome notations in what follows. 
Being symmetrical, this tensor can be diagonalised, with the principal strain components denoted as $\e_{(1)}$, $\e_{(2)}$, $\e_{(3)}$ \citep{Landau1959}. In a body subjected to deformation we can write the new coordinates $\hbfr$ as:
\begin{equation}\label{Eq:HatDef}
	\hbfr = (\x(1+\e_{(1)} ),\y(1+\e_{(2)} ),\z(1+\e_{(3)} ).
\end{equation}
The deformed volume and area of a strained body, $ \hV$ and $\hA$, can be described in the same way as functions of the strain in each direction. It can be shown that $ \hV =  V(1+\e_{11} + \e_{22} + \e_{33}) =  V(1 + \e_{(1)} + \e_{(2)} + \e_{(3)})$, using the invariance property of the trace of the matrix, and, for a surface perpendicular to the $\z$ direction: $ \hA = A(1+\e_{11} + \e_{22}) = A(1 + \e_{(1)} + \e_{(2)})$  (see Assumptions \ref{hyp:flat}, \ref{hyp:eq}-\ref{hyp:me}) \citep{Landau1959}. We want to highlight here that the previous equivalence of the strained area ($\hA$) in two references: $i$) with axis $\x$ and $\y$ in a generic orientation, $ii$) with axis oriented according to the direction of the principal strain components, is valid under the assumption that the surface is perpendicular to the third direction $\z$. If that is not the case the relations must be modified accordingly by considering the two directions parallel to the surface.
According to this notation, we can write the infinitesimal variation of volume, $\de V$, as 
 $\de V = (\hV - V) =  V(\e_{(1)} + \e_{(2)} + \e_{(3)})$. The same applies for the definition of the  infinitesimal variation of the area, $\de A$.

\section{Liquid Interface} \label{Sec:LiqV}

\subsection{Liquid-Vacuum Interface}
\noindent If we put a liquid in a closed container in contact with vacuum, we will observe some of the molecules moving from the liquid bulk into the vacuum space. This vapour phase will have a definite pressure that depends on the temperature and composition of the liquid. If the vapour pressure is small enough to be negligible, we can assume that the liquid is in contact with vacuum. The Liquid-Vacuum (LV) interface represents the simplest case to consider.  We will start our derivation from here to develop all the concepts we will need in the rest of the work. Some of the concepts derived for the LV interface system are self-evident, but we think a full derivation will be helpful for later sections, when more complicated systems are introduced.

Let us consider a liquid phase in contact with vacuum, i.e. that has a surface exposed to vacuum.  By using the first and second law of thermodynamic we can write:
\begin{equation}\label{Eq:ensurf}
	\de U = T \de S  + \mu \de N +  \de \work
\end{equation}
where $U$ is the internal energy, $T$ is the temperature, $S$ is the entropy, $\mu$ is the chemical potential, and $N$ is the number of particles in the system. The last term, $\de \work$, represents the reversible work done on the system. In the bulk it consists of the mechanical work, $\de \work^M$, done by the change of the system volume, $V$, at pressure $P$, i.e. $\de \work^M = -P\de V$. When the system contains an interface, $\de \work$ also includes the work required to create such an interface, $\de \work^A$, i.e. $\de \work = \de \work^A + \de \work^M$. 
The reason why we prefer to include the volume work done by the system and the work needed to create the interface into a single work term will be clear in \cref{Sec:MolExpl} when the stress tensor is introduced. \\
Since the pressure $P = 0$ in the liquid-vacuum system, the mechanical work is $\de \work^M = 0$.  Thus the total work reduces to $\de \work = \de \work^A$. 

\subsection{Stress tensor and the origin of Surface Free Energy} \label{Sec:MolExpl}

\noindent The local environment seen by an atom in the bulk of a condensed phase is different from the environment seen by the same atom on the surface \citep{Brown1974}. In the former case, each atom is fully surrounded by other atoms and in the latter case, an atom is surrounded by other atoms only on one side, the other being exposed to the vacuum. This difference between bulk and surface has a strong effect on the total potential energy associated to each atom. Liquid phase exists because of the long-range attraction between atoms, which results in the tendency of the atoms to congregate into liquid clusters (droplets). This leads to the net attraction of the atoms on the surface towards the bulk  and to a higher potential energy of the surface atoms relative to the bulk atoms. From these effects derives the tendency of the surface to shrink (as described in \cref{Sec:intro}).\\
In creating a new interface we need therefore to perform some work on the system. 
If the formation of the new surface is carried out at constant pressure and temperature, then the relevant thermodynamic quantity is the Gibbs free energy ($\boldsymbol{\mathcal{G}}$), while for the surface formation at constant volume and temperature, the Helmholtz free energy ($\HFE$) should be used \citep{Cahn1979,Haiss2001}.

The difference in free energy between the system with and without a surface can be therefore related to the reversible work needed to separate the atoms in the bulk in order to create an exposed surface, i.e. it is part of the work term $\de \work$ in \cref{Eq:ensurf}. \citet{Eriksson1969} introduced the concept of {\em cleaving}, as the reversible transformation that creates an interface in a sample, but we will postpone its exact definition and the related discussion to a later section. Here, we want just to consider the concept of cleaving as given, and as we can intuitively expect, the infinitesimal work $\de \mathcal{W}^A$ needed to cleave a sample in order to create an exposed surface of size $\de A$, is proportional to the area created. We will call $\gamma$ the proportionality constant such that  \cite{Ono1960}
\begin{equation}\label{Eq:sfe}
	\de \work^A=\gamma \de A. 
\end{equation}

Another way to describe $\de \work$, which proves to be more general and can be easily extended to different systems, is represented by the stress tensor formalism.  

When a material is deformed, it resists the deformation. This resistance is represented by internal forces that oppose such a deformation, which can be described using the concept of stress. The state of stress of the material is represented by the Cauchy stress tensor $\stress(\bfx)$, which is a three dimensional rank-2 tensor, function of the position $\bfx$, and is usually written in matrix notation \citep{Landau1959}. In this work we will use lower case Greek letters to indicate the component of the stress matrix: $\s_{ij}$ with $i,j=1,2,3$. The stress in a material is closely related to the notion of pressure. In particular, the pressure tensor $\Ptens$ can be defined from the stress tensor as the tensor (in matrix notation) having every entry equal to the negative of the stress tensor:  $\p_{ij}=-\s_{ij}$ with $i,j=1,2,3$. This convention stems from the fact that pressure is a positive quantity which always represents a compressive stress, while the convention for stresses says that a compressive stress is negative. 

\begin{figure}[h]
	\begin{center}
			\includegraphics[scale=0.8]{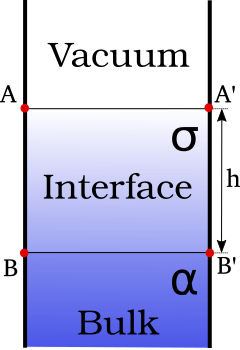}
	\end{center}
	\caption{Sketch of the ideal system composed of a liquid phase with a surface exposed to vacuum}
	\label{Fig:Vacsyst}
\end{figure}

The Cauchy stress tensor for fluid phases assumes a particularly simple form, thanks to the following assumption:
\begin{enumerate}[label=\roman*]
\setcounter{enumi}{5}
	\item The bulk of the fluid phase is isotropic and homogenous; \label{Hyp:Hom}
\end{enumerate}
With this assumption, the stress tensor in fluid phases, in matrix components, reduces to $\stress=\mbox{diag}(-P,-P,-P)$, where $P$ is the pressure in the fluid phase (for a derivation see Section I of the Supplemental Material (SM)).
However, this diagonal form of the stress tensor is valid within the bulk of the system only. The presence of the surface modifies $\stress$ \cite{Kirkwood1949,Gurney1947} and the stress tensor in the LV system is no longer given by: $\stress=\mbox{diag}(-P,-P,-P)$. This latter fact makes this theory difficult to use in practice, unless we know the actual form of $\stress$. However, the perturbations caused by the surface on the whole system decay within a few molecular diameters from the interface \citep{Gutman1995,Gurney1947}. The physical situation we have is that $\stress$ is diagonal with $-P$ on the main diagonal in the whole system, except a very narrow volume close to the interface. 
This picture suggests a different way to describe the stress tensor. We divide our system into a \textit{bulk} region, $\alpha$, which does not feel the presence of the interface, and an intermediate region, $\sigma$, where the properties are modified by the interface, as shown in \cref{Fig:Vacsyst}.  
The different regions are separated by two planes $AA'$ and $BB'$ parallel to the (flat) interface. The extent of the intermediate region $\sigma$ is given by its height $h$. The strength of this approach, as it will be shown later, is that the final results do not depend on $h$, as long as all the perturbations from the presence of the interface are confined within $\sigma$.  If the system depicted in \cref{Fig:Vacsyst} is in thermodynamic equilibrium, then intensive quantities do not change between the regions.  That is, given an intensive quantity $X$ (e.g., temperature), $X^\alpha(BB') = X^\sigma(BB') = X$, and analogously for plane $AA'$. 
This approach was proposed first by \citet{Guggenheim1940}, and compared to Gibbs's dividing surface formalism has the advantage of being less abstract by recognising that, on a microscopic level, the properties of the interface originate from a region of finite thickness $\sigma$. \\
Thus, we can define thermodynamic quantities for each of the different finite-thickness regions into which the system is divided:
\begin{align}
	\de U^{\alpha} & = T \de S^{\alpha}  + \mu \de N^{\alpha} - P\de V^{\alpha} \\
	\de U^{\sigma} & = T \de S^{\sigma}  + \mu \de N^{\sigma} + \de \work^{\sigma} 
\end{align}
where $\de \work^\sigma$ is the mechanical work in the region $\sigma$ as described in \cref{Eq:ensurf}. Note that only the thermodynamic equation relative to interface region $\sigma$ contains the generic work term $\de \work^\sigma$. The bulk region contains only the mechanical work term $- P\de V^{\alpha}$.

To identify different spatial locations in the system depicted in \cref{Fig:Vacsyst}, we introduce a coordinate system $x$-$y$-$z$ with the $z$-axis perpendicular to the planes $AA'$ and $BB'$.  The mechanical equilibrium and the constancy of intensive quantities implies that the $z$ component of the stress tensor is equal to zero in $\sigma$ and $\alpha$. It follows that from the isotropy of the bulk liquid phase the stress tensor in $\alpha$, $\stress^\alpha$, is the null tensor.  

In the region $\sigma$ the stress tensor, $\stress^\sigma$, will be in general different from $\stress^\alpha$ because of the presence of the interface. This qualitative statement is more subtle than it looks. Since we assumed the continuity of the physical properties on the plane $BB'$, we should have $\stress^\alpha=\stress^\alpha(BB')=\stress^\sigma(BB')$. This means that the stress tensor in $\sigma$ is not constant within $\sigma$. The latter fact means, in turn, that assumption \ref{Hyp:Hom} is not valid in $\sigma$. The only statement we can make about $\sigma$ is a weaker symmetry property than the ones valid for $\alpha$, replacing assumption \ref{Hyp:Hom} valid for the bulk \cite{Rowlinson2013} with:
\begin{enumerate}[label=\roman*]
\setcounter{enumi}{5}
	\item$'$ $\sigma$ has a circular symmetry in the planes parallel to the interface, i.e. rotations around the $\z$-axis leave properties unchanged. \label{Hyp:Hom2}
\end{enumerate}

From assumption \ref{Hyp:Hom2}$'$ it follows that $\stress^\sigma(\bfx)$ must be a diagonal tensor $\stress^\sigma(\bfx)=\mbox{diag}(\s^\sigma_{11}(\bfx),\s^\sigma_{22}(\bfx),\s^\sigma_{33}(\bfx))$ with $\eta^\sigma_{11}(\bfr) = \eta^\sigma_{22}(\bfr)$. 
Now we impose the mechanical equilibrium condition, which can be written in terms of the divergence of the stress tensor \citep{Landau1959} $\nabla \cdot \stress^\sigma=\mathbf{0}$, where $\mathbf{0}$ is the three-dimensional zero vector. This latter condition translates into
\begin{align}\label{Eq:mechEq}
	\Dpartial{\s^\sigma_{11}(\bfx)}{x}  = \Dpartial{\s^\sigma_{22}(\bfx)}{y}  = \Dpartial{\s^\sigma_{33}(\bfx)}{z} = 0,
\end{align}
from which it follows that the first two components of the stress tensor are function only of $\x$, $\s^\sigma_{11}(z)=\s^\sigma_{22}(z)$ and the third component is constant, $\s^\sigma_{33}=const$. From the fact that on the plane $AA'$ (see \cref{Fig:Vacsyst}) $\s^\sigma_{33}$ must be zero because the system is in contact with vacuum, it results that $\s^{\sigma}_{33} = 0$ everywhere in $\sigma$.

The tensor $\stress^\sigma(z)$ can be thought of as the sum of two contributions, one representing the average stress that would be in the region $\sigma$ if the interface was not there, the second representing the excess stress relative to the bulk due to the presence of the interface:
\begin{align}\label{Eq:Excs}
	\stress^\sigma(z) & = \cip{\stress^\sigma(z) - \stress^\alpha} +\stress^\alpha =\exstr(z) +  \stress^\alpha.
\end{align}
The excess stress tensor, $\exstr^\sigma$, is defined by the last equality in \cref{Eq:Excs} and can be written as:
\begin{equation}
	 	\exstr(z) =  \begin{bmatrix}
\cex_{11}(\bfz)& 0 & 0 \\ 
0 &  \cex_{22}(\bfx)  & 0 \\
0 & 0 & \cex_{33}(\bfx)
\end{bmatrix} =  \begin{bmatrix}
\s_{11}^\sigma(\bfx) + P& 0 & 0 \\ 
0 &  \s_{22}^\sigma(\bfx) + P & 0 \\
0 & 0 & 0
\end{bmatrix} = \begin{bmatrix}
-\p_{11}(z)  & 0 & 0 \\ 
0 & -\p_{22}(z) & 0 \\
0 & 0 & 0
\end{bmatrix}
\end{equation}
where $\p_{11}(z) = \p_{22}(z)$ because $\eta^\sigma_{11}(\bfx) = \eta^\sigma_{22}(\bfx)$. 

The work associated within the system by the excess stress tensor $\mathcal{E}^\sigma(z)$ can be written by considering the strain tensor representing the infinitesimal displacements $\e_{ij}$ compatible with the constraints of the system\footnote{In this case the constraints are that there cannot be any shear in the liquid, so all the off-diagonal elements of the strain tensor are zero.} (see end of the \cref{Sec:intro}):
\begin{equation}\label{Eq:strain}
	\strain = \begin{bmatrix}
\e_{11} & \e_{12}  & \e_{13}  \\ 
\e_{21} & \e_{22}  & \e_{23}  \\
\e_{31} & \e_{32}  & \e_{33}
\end{bmatrix} 
\end{equation}
However, all the strains must be considered small enough not to cause plastic deformation in the material, i.e. the theory developed here can be applied only to elastic deformations. A generalisation of this theory for finite strains will be considered in the future.  The total mechanical work $\de \mathcal{W}$ can be written as (summation over repeated indexes implied, see Section S.3.A of the SM):
\begin{align}\label{Eq:work}
	\de \mathcal{W} = \int\int\int_{V^\sigma}{\cip{\cex_{ij}(z)\e_{ij}}} \de x \de y \de z  = \de \mathcal{W}^A 
\end{align}
where in this case $\de \mathcal{W}^M=0$.
Since  the argument  of the integral in \cref{Eq:work} depends only from $z$ we can directly integrate over the two directions $x$ and $y$ obtaining the area $A$. We can therefore write (see Section S.3.B of the SM):
\begin{align}\label{Eq:workstress}
	\de \mathcal{W}^A  = A f_{11}\e_{11} + A f_{22}\e_{22}  
\end{align}
where $A$ is the area of the surface and we introduced a degenerate two-dimensional, second order, surface stress tensor, $\surfst$, which reads:
\begin{equation}\label{Def:surfstLV}
	\surfst = \begin{bmatrix}
\f_{11} & \f_{12}  \\ 
\f_{21} & \f_{22}   \\ 
\end{bmatrix} = \begin{bmatrix}
-\int_{-\infty}^{+\infty}{ \de z \,\, \p_{11}(z) } & 0  \\ 
0 & -\int_{-\infty}^{+\infty}{ \de z \,\,\p_{22}(z) }   \\ 
\end{bmatrix} 
\end{equation}
The surface stress tensor was introduced by \citet{Herring1951b} as a generalisation of the Shuttleworth equation to non-symmetric surfaces. In our derivation, it represents the most important quantity describing all the information we need for the surface properties. Its meaning was disputed in literature \citep{Gutman2014}, and for this reason we will give a detailed statistical mechanical formulation of this quantity and the Shuttleworth equation in the next sections. 

In \cref{Def:surfstLV} we can replace the integration boundary with $\pm\infty$ by noting that, by the definition of the excess quantities, the integrands in \cref{Def:surfstLV} are equal to zero outside the interfacial region $\sigma$.  Therefore, the value of the integral does not depend on the integration interval, as long as it contains the whole of the region where properties deviate from those in $\alpha$ or in the vacuum. 

For the system considered here, $\p_{11}(z) = \p_{22}(z) = \p(z)$  (see disscussion of Assumption \cref{Hyp:Hom2}$^\prime$), and the SS tensor reduces to a single scalar, which we call $\fs$. We can therefore rewrite \cref{Eq:workstress} as
%
\begin{equation}\label{Eq:finworkLV}
	\de \mathcal{W}^A = \sqp{-\int_{-\infty}^{+\infty}{ \de z\cip{\p(z)}}}\de A
\end{equation}
where we use the fact that the area change due to strain is given by $\de A = A(\e_{11} + \e_{22})$.

We have presented two different interpretations of the work $\de \mathcal{W}^A$, a free energy as the work to create a new interface and a stress coming from the fact that an interface was created, and we concluded with two different expressions describing the same quantity. 
However, comparing \cref{Eq:finworkLV} and \cref{Eq:sfe} we can write: 
\begin{equation}\label{Eq:hydrStressLV}
\fs = \gamma = \int_{-\infty}^{+\infty}{ \de z\cip{\p(z) }} 
\end{equation}
which represents the connection between the mechanical and thermodynamics interpretation of the interfacial work. 
This identification can be made more rigorous by using the statistical mechanics description of the system and will be presented in \cref{Sec:Shuttl}.


\subsection{Liquid-Vapour Interface}\label{Sec:Liq-vap}

\begin{figure}[h]
	\begin{center}
			\includegraphics[scale=0.8]{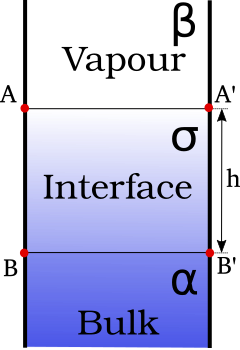}
	\end{center}
	\caption{Sketch of the ideal system composed by a phase with a surface exposed to vacuum}
	\label{Fig:Vaposyst}
\end{figure}

\noindent Let us consider a more realistic system, where a liquid is in contact with its vapour. The main difference with the system described in previous section is that $P\neq0$. Now, we can identify the liquid bulk $\alpha$, the interface region $\sigma$, and the vapour bulk $\beta$, as sketched in \cref{Fig:Vaposyst}.

The thermodynamic equations for this system become:
\begin{align}
	\de U^{\alpha} & = T \de S^{\alpha}  + \mu \de N^{\alpha} - P\de V^{\alpha} \\  
	\de U^{\sigma} & = T \de S^{\sigma}  + \mu \de N^{\sigma} + \de \work^{\sigma}  \\
	\de U^{\beta}  & = T \de S^{\beta}   +  \mu \de N^{\beta}  - P\de V^{\beta}  
\end{align}
The stress tensor in $\alpha$ is equal to stress tensor in $\beta$, $\stress^\alpha=\stress^\beta$, while in $\sigma$ it becomes:
\begin{equation}
 	\stress^\sigma(z) =  \begin{bmatrix}
-\p_{11}(z) & 0  & 0 \\ 
0  & -\p_{22}(z) & 0 \\
0  & 0  & -P 
\end{bmatrix}
\end{equation}
from which the excess stress tensor, ($\exstr(z)=\stress^\sigma(z)-\stress^\alpha$) can be obtained.
The total mechanical work in $\sigma$, $\de \work$ is given, using the strain tensor defined in \cref{Eq:strain}, by (see Section S.3.C of the SM):
\begin{align}\label{Eq:work:LL}
	\de \mathcal{W} & = \de A \int_{-\infty}^{+\infty}{ \de z\cip{P-\p(z)}} - P\de V^\sigma =\de \mathcal{W}^A + \de \mathcal{W}^M
\end{align}
where we used the fact that the change of volume due to strain is $\de V = V(\e_{11}+\e_{22}+\e_{33})$ and the fact that $\p_{11}(z)=\p_{22}(z)=\p(z)$. The surface stress is given by:
\begin{equation}\label{Def:surfstLVap}
		\surfst = \begin{bmatrix}
		 \int_{-\infty}^{+\infty}{ \de z\cip{P-\p(z)}}  & 0  \\ 
		0 &  \int_{-\infty}^{+\infty}{ \de z\cip{P-\p(z)}}   \\ 		
\end{bmatrix} 
\end{equation}
and the surface free energy, $\gamma$, is:
\begin{equation}\label{Eq:gammaLVap}
	\gamma = \int_{-\infty}^{+\infty}{ \de z\cip{P-\p(z)}}. 
\end{equation} 
\Cref{Eq:gammaLVap} is the equation for $\gamma$ usually known in the literature as the ``mechanical definition'' of the SFE \cite{Rowlinson2013}. The usual interpretation of the mechanical definition of $\gamma$ is a difference between the normal component of the stress ($P$ in our case) and half of the tangential component of the stress (i.e. $\frac12\cip{\p_{11}(z)+\p_{22}(z)} = \pi(z)$ in our case). In our derivation there is no normal contribution to the interfacial properties, i.e. the excess quantities which represent the interface do not depend on any normal contribution. The presence of $P$ in \cref{Eq:gammaLVap} results from our definition of the excess stress tensor as the difference between the stress tensor in $\sigma$, which contains the contributions given by the presence of the interface, and the pressure. 
The fact that the difference between the two tensors returns an expression where the pressure $P$ is present does not depend by the presence of normal contributions to the interface stress, but on the hypothesis of hydrostatic stress in the liquid which means that all the diagonal components of the stress in the bulk phases $\alpha$ and $\beta$ are equal to the pressure $P$ (i.e. $\p^\alpha_{11}=\p^\alpha_{22}=\p^\beta_{11}=\p^\beta_{22}=P$, see Assumption \ref{Hyp:Hom}). The normal component of the stress tensor $\p_{33}(z)$ does not enter into the definition of $\gamma$. 
For liquids this remark is not essential, but for solids becomes crucial  since the bulk internal stress can be different in different directions.


\section{Solid interface}\label{Sec:SolInt}

\noindent When at least one of the two phases is solid, there are two complications that we need to add to the previous framework. Solids can resist to shear stress, meaning that the stress tensor can have off-diagonal components and, more importantly, solids can be in a {\em non-hydrostatic} state of stress. If the bulk solid is hydrostatically stressed, then the reasoning that brought us to the equations for the SS tensor for the case of liquids is still valid (see \cref{Def:surfstLV} for solid-vacuum and \cref{Def:surfstLVap} for solid in contact with its vapour or liquid) with the added complexity given by the presence of the off-diagonal terms. However, the main difference between liquid and solid interfaces resides in the fact that in the latter case there is no more equality between the SS and SFE (i.e. \cref{Eq:gammaLVap} does not hold for solids). These two quantities are different and their relationship is given by the Shuttleworth equation, which will be discussed in the next Section (see \cref{Sec:Shuttl}).

In this Section we will analyse the solid-vapour (or solid-liquid) interface. The region $\alpha$ represents the bulk of the solid, the region $\beta$ the bulk of the fluid phase, and as before, we call $\sigma$ the interface region.  

The thermodynamics equations for solid-vapour interface, which represent the most general equations for a single-component system are given by:
\begin{align}
	\de U^{\beta} & = T \de S^{\beta} + \mu \de N^{\beta}  -P \de V^\beta   \\
	\de U^{\sigma} & = T \de S^{\sigma}  +  \mu \de N^{\sigma} + \de \work \label{Eq:thermoSsol} \\
	\de U^{\alpha} & = T \de S^{\alpha}  +  \mu \de N^{\alpha} + V^\alpha(\stress^\alpha: \e^\alpha),
\end{align}
where 

\begin{equation}\label{Eq:HalphaS}
	\stress^\alpha = \begin{bmatrix}
 \s_{(1)}^\alpha & 0 & 0 \\ 
0  &  \s_{(2)}^\alpha & 0 \\
0 & 0 & \s_{(3)}^\alpha
\end{bmatrix} = \begin{bmatrix}
 -P & 0 & 0 \\ 
0  &  -P & 0 \\
0 & 0 & -P
\end{bmatrix}.
\end{equation}
where we assumed, without loss of generality, that the system is oriented along the direction of the principal stresses.
The chemical potential $\mu$ that appears in the above equations is the chemical potential of the component in the fluid phase, as derived by Gibbs, which described the equilibrium conditions for this kind of systems \cite{Gibbs1957,Frolov2010,Frolov2010b}. We will always use the symbol $\mu$ with this meaning when a solid interface is considered.
The symbol $:$ represents a dyadic product, and it is equal to $(\stress^\alpha: \e^\alpha) = \s_{ij}^\alpha\e^{\alpha}_{ij}$, where the Einstein summation convention is implied.
The form of the tensor $\stress^\alpha$ in \cref{Eq:HalphaS} is consequence of Assumptions ii.\ref{hyp:me}-\ref{hyp:flat}.0.
Being real and symmetric, the stress tensor $\stress^\alpha$ tensor can be put in diagonal form, with its three eigenvalues on the main diagonal. These eigenvalues are called \textit{principal stresses} \cite{Nye1985} and we will indicate them as $\s^\alpha_{(i)}$, $i=1,2,3$. Directions along which the stress tensor is diagonal are called \textit{principal directions}.  
As a consequence of mechanical equilibrium hypothesis (see Assumption ii.\ref{hyp:me}), in a system with an interface one of the axes is always aligned with the direction of one of the principal stresses along the normal to the interface. Here, we assume that all the axes are oriented along the principal stress directions, from which $\s^\alpha_{ij} = -\p^\alpha_{ij} = \delta_{ij} \s_{(i)}^\alpha$.

\subsection{Hydrostatically stressed solid}

If the solid is in a hydrostatic state of stress then $\p_{ij}^\alpha = \delta_{ij} P$, where $P$ is the pressure of the phase (liquid or vapour) in contact with the solid \citep{Sekerka2004}. The derivation of the surface stress tensor will follow steps similar to the one presented for liquids (see \cref{Eq:work:LL}), with some added complications given by two essential differences of the stress tensor in the interface region $\stress^\sigma(z)$ when solids are present. Firstly, we cannot assume that $\s^\sigma_{11}(z)=\s^\sigma_{22}(z)$ as we did for the liquids (see discussion of \cref{Eq:mechEq}). The Assumption \ref{Hyp:Hom2}$^\prime$, in general, does not work for solids. We will give an example of this in the results Sections (see \cref{Sec:Results}), where it is shown that the orientation (110) of face-centered cubic crystal gives different stresses for different directions.  Secondly, even if the solid is oriented along the direction of the principal stresses, which ensures that the stress tensor in the solid bulk ($\alpha$) has the negative of the liquid pressure on the main diagonal, this is not true for the interface region any more. The presence of the interface gives rise to additional stresses, as explained for the liquid case, with the only difference that solids now allow the presence of additional stresses in the non-diagonal terms of the stress tensor \citep{Nozieres1988}. Therefore, we can write for the stress tensor in the interface region $\sigma$:
\begin{equation}\label{Eq:HsigmaS}
	\stress^\sigma(z) = \begin{bmatrix}
 \s_{11}^\sigma(z) & \s_{12}^\sigma(z) & 0 \\ 
 \s_{21}^\sigma(z)  &  \s_{22}^\sigma(z) & 0 \\
0 & 0 & \s_{33}^\sigma(z)
\end{bmatrix} = \begin{bmatrix}
 -\p_{11}^\sigma(z) & -\p_{12}^\sigma(z) & 0 \\ 
 -\p_{21}^\sigma(z)  &  -\p_{22}^\sigma(z) & 0 \\
0 & 0 & -P
\end{bmatrix}.
\end{equation}
from which the excess stress tensor becomes:
\begin{equation}
	 	\exstr(z) =  \begin{bmatrix}
-\p^\sigma_{11}(z) + P & -\p_{12}^\sigma(z)  & 0 \\ 
-\p_{21}^\sigma(z) &  -\p^\sigma_{22}(z) + P & 0 \\
0 & 0 & 0
\end{bmatrix}
\end{equation}
The total mechanical work, $\de \work$, in $\sigma$ is given by:
\begin{align}\label{Eq:work:SS}
	\de \mathcal{W} & = A\int_{-\infty}^{+\infty}{ \de z\cip{ P-\p_{11}(z) }\e_{11}} + A\int_{-\infty}^{+\infty}{ \de z\cip{ P-\p_{22}(z)}\e_{22} } + A\int_{-\infty}^{+\infty}{ \de z\cip{-\p_{12}(z) }\e_{12}} \nonumber \\
		&  + A\int_{-\infty}^{+\infty}{ \de z\cip{ -\p_{21}(z)}\e_{21} } - PV^\sigma(\e_{11}+\e_{11}+\e_{33})  =\de \mathcal{W}^A + \de \mathcal{W}^M  
\end{align}
where we put again together the terms related to the mechanical work and those related to the interface. The surface stress tensor assumes the known expression \citep{Ibach1997}: 
\begin{equation}\label{Eq:SSsolid}
		\surfst = \begin{bmatrix}
 \int_{-\infty}^{+\infty}{ \de z\cip{P-\p_{11}(z)}}  & \int_{-\infty}^{+\infty}{ \de z\cip{-\p_{12}(z)}}   \\ 
\int_{-\infty}^{+\infty}{ \de z\cip{-\p_{21}(z)}}  &  \int_{-\infty}^{+\infty}{ \de z\cip{P-\p_{22}(z)}}   \\ 
\end{bmatrix} 
\end{equation}
We want to highlight here that \cref{Eq:gammaLVap} is not true in general for solids. For solids, each component of the surface stress tensor can vary independently, and the connection with the surface free energy is given by the Shuttleworth equation, as we will show in \cref{Sec:Shuttl}. This latter fact has great importance in the calculation of the properties of solids. While for liquids, obtaining the stress tensor from the simulation is equivalent to calculating $\gamma$ using, e.g. the cleaving approach, for solids it is not true in general, and the calculation method must be decided with respect the property we are interested in.

\section{Relationship between SS and SFE}\label{Sec:Shuttl}

\noindent In previous sections we have derived the work term $\de \work^A$ for liquid and solid systems. We showed that for the formers this work can be described by a scalar quantity, $f^S$, whereas in the latter case we obtained a tensorial quantity, $\surfst$.

Let us now summarise the derivation we presented in previous Sections.

For a single-component bulk phase (e.g. $\beta$), the most general variation of Helmholtz free energy, $\HFE$, is given by:
\begin{equation}
	\de \HFE^\beta = -S^\beta\de T + V^\beta(H^\beta: \strain^\beta)+\mu \de N^\beta
\end{equation}
whereas, for the surface layer $\sigma$ we have (see \cref{Eq:thermoSsol,Eq:work:SS}):
\begin{align} \label{Eq:FreeEn}
	\de \HFE^\sigma & = -S^\sigma\de T + V^\sigma(H^\beta: \strain^\sigma)+\mu \de N^\sigma + A(\surfst:\strain^\sigma) \nonumber \\
	& = -S^\sigma\de T + \sum_{i=1}^{3}V^\sigma\s_{(i)}^\beta \e^\sigma_{ii}+\mu \de N^\sigma + \sum_{i=1}^{2}\sum_{j=1}^{2}A\f_{ij} \e^\sigma_{ij}
\end{align}
where we used the surface stress tensor obtained in \cref{Eq:SSsolid} and we included explicit sum sign to emphasize the different dimension of the surface stress tensor and the stress tensor $\stress^\beta$ (note that we are in the hypothesis that $\stress^\beta=\stress^\alpha$). In the Assumptions described in \cref{Sec:intro}, \cref{Eq:FreeEn} is valid for both solid-liquid and liquid-liquid interfaces.

If we now specialise \cref{Eq:FreeEn} to the case where both $\alpha$ and $\beta$ are fluid phases, we have $\s^\beta_{(i)}=\s^\alpha_{(i)}=-P$, with $i=1,2,3$, and $f_{11}=f_{22} = \fs$ and $f_{12}=f_{12}=0$ (i.e. the tensor $\surfst$ assumes the form shown in \cref{Def:surfstLV}). We therefore can write:
\begin{align} \label{Eq:FreeEnLiq}
	\de \HFE^{\sigma} & =  -S^\sigma\de T - V^\sigma(\e_{11}+\e_{22}+\e_{33}) P + A(\e_{11}+\e_{22}) \gamma  + \mu \de N \nonumber \\
	& = -S^\sigma\de T- P \de V^\sigma + \gamma \de A + \mu \de N
\end{align}
as expected \citep{Marchenko1980}, where we replaced the symbol for the scalar component of the SS, $\fs$, with $\gamma$, using the identification we discussed in \cref{Eq:finworkLV,Eq:hydrStressLV}. Therefore, we can write for a transformation at constant temperature, volume and number of atoms:
\begin{equation}\label{Eq:SFEL}
 	\gamma =  \cip{\pard{ \HFE^\sigma}{A}}_{N,V,T} 
\end{equation}
A transformation of this kind can be performed within MD simulations \citep{Davidchack2003}. 
This transformation can be carried out continuously and reversibly (with certain precautions) within the framework of the thermodynamic integration and will be analysed in detail in a later section.
 
For interfaces involving solids, we cannot simplify the equations as before (see \cref{Eq:FreeEnLiq}). In this case we have to talk about a transformation at constant temperature, number of atoms, and strain in a particular direction, i.e. we can allow the variation of the strain in only one direction at a time. If we choose the $\x$ (or $x$) direction, we obtain from \cref{Eq:FreeEn}:
\begin{align}\label{Eq:Fsol}
  \cip{V^\sigma\s_{(1)}^\beta + A\f_{11}} = \cip{\pard{ \HFE^\sigma}{\e_{11}}}_{N,T,\e_{12},\e_{21},\e_{22},\e_{33}} 
\end{align}
The form of the stress term reported in the previous equations $\cip{V^\sigma\s_{(1)}^\beta + A\f_{11}}$ does not look like the usual term encountered in thermodynamics, which are in the form $X\de Y$, with $X$ and $Y$ being a pair of conjugate variables. This apparent discrepancy could have deep consequences for the theory developed here. \citet{Hermann1973} described the mathematical structure of the thermodynamics, which prescribes that the quantities involved in a transformation appear in the form of $X\de Y$. Arguments based on Hermann's theory were used to criticise \citep{Bottomley2009,Marichev2009} the Shuttleworth equation and its consequences in different works \citep{Muller2004,Ibach1997,Rusanov2005}, with much subsequent debate about the meaning and validity of this equation \citep{Bottomley2009b,Bottomley2009c,Bottomley2009d,Bottomley2010,Bottomley2010b,
Eriksson2009,Eriksson2010,Eriksson2010b,Eriksson2011,Ibach2009,Hecquet2010,
Gutman2011,Gutman2011b}. 

However, in our view, there is no discrepancy between the equations derived here and Hermann's theory. In \cref{Eq:Fsol} the quantity  $\cip{V^\sigma\s_{(1)}^\beta + A\f_{11}}$ is just a different way to write the component $\s_{11}^\sigma(z)$ of the stress tensor, $\stress^\sigma(z)$ in $\sigma$ (see \cref{Eq:Excs}). The original term is $\s_{11}^\sigma(z) \e_{11}$  which then appears in the correct form $X \de Y$.

In this case the stress tensor $f_{11}$ is not just equal to the variation of the free energy because of the presence of the term proportional to the volume of phase $\sigma$. However, the term $V^\sigma\s_{(1)}^\beta$ has a clear interpretation as the stress in the region $\sigma$ if there is no interface. From the definition of the Helmholtz free energy in the bulk, for a transformation at constant temperature, number of atoms and variation of the dimension in directions $y$ and $z$ (i.e. the same constraints we used for \cref{Eq:Fsol}) we can write:
\begin{equation}
	  V^\sigma\s_{(1)}^\beta = \cip{\pard{ \HFE^\sigma_{bulk}}{\e_{11}}}_{N,T,\e_{12},\e_{21},\e_{22},\e_{33}}
\end{equation}
where $\HFE^\sigma_{bulk}$ represents the variation of the Helmholtz free energy in the region $\sigma$ at constant temperature, volume, number of molecules and strain in $y$ and $z$ direction, when the region $\sigma$ does not contain a surface (bulk). By replacing the previous equation in \cref{Eq:Fsol}, we eventually obtain:
%
\begin{equation}\label{Eq:fFreeE}
	A\f_{11} = \cip{\pard{ }{\e_{11}} \cip{\HFE^\sigma - \HFE^\sigma_{bulk}}}_{N,T,\e_{12},\e_{21},\e_{22},\e_{33}}  = \cip{\pard{\HFE^\sigma_{exc} }{\e_{11}}}_{N,T,\e_{12},\e_{21},\e_{22},\e_{33}} 
\end{equation}
where we introduced the excess of Helmholtz Free Energy, $\HFE^\sigma_{exc}$. The concept of excess Helmholtz Free Energy is not new in the discussion of properties of interfaces\citep{Johnson1959,Linford1978}. However, it is important to highlight here the fact that we are using this terminology with a different meaning with respect to previous works, namely to indicate the difference in free energy between systems with and without an interface.

We have therefore come to the conclusion that, while for liquids interface description can be reduced to a single scalar quantity, $\fs$, whereas for solids the interface is described by the tensorial quantity $\surfst$.

Last steps of our derivation involve showing two facts: $i$) $\fs$ can be indeed identified with $\gamma$, i.e. the Surface Free Energy which was defined in \cref{Eq:sfe}, $ii$) the SFE, $\gamma$ is related to the SS and this relation is represented by the {\em Shuttleworth equation}  \citep{Shuttleworth1950}:
\begin{equation} \label{Def:Shutt}
	f_{ij} = \delta_{ij}\gamma + \Dpartial{\gamma}{\e_{ij}} .
\end{equation}	
where $\delta_{ij}$ is the Kronecker delta. 

Note that if we assume the validity of the point $ii$ (i.e. we assume the Shuttleworth equation is correct) then point $i$ automatically follows. In fact, for liquids, the derivative of the SFE with respect the strain is zero because liquids cannot be stretched \citep{Couchman1972,Couchman1973} and thanks to this $f_{11}=f_{22}=\fs=\gamma$ and $f_{12}=f_{21}=0$. Therefore, in following sections we will focus on the Shuttleworth equation only.

In \citep{Couchman1972,Couchman1973,Couchman1976}, the thermodynamic transformation which creates a new area in a liquid system is called ``plastic deformation''.  However, we prefer to follow the prescription of \citet{Kramer2007}, where it is pointed out that the accepted definition of ``plastic transformation'' refers to non-reversible transformations only, contrary to the assumptions in this work. In order to avoid confusion, we will simply say that liquids cannot be elastically stretched, having in mind the discussion in \citep{Couchman1972,Couchman1973}.

We want to highlight here that \cref{Eq:fFreeE} represents the crucial link between the classical thermodynamics and the microscopic statistical mechanical description.
In next Sections, we will obtain the Shuttleworth equation from the concepts of the statistical-mechanical description of the interface, on the basis of interatomic interactions within atoms.
The Shuttleworth equation was at the centre of different debates in literature which dates back to 1995. \citet{Gutman1995} argued that this equation is incorrect on the basis that it relates two incompatible thermodynamic mechanisms for the creation of the surface  (cleaving and stretching) pertaining to two different systems (liquids and solids). However, as we will show later, when the cleaving mechanism will be introduced, the cleaving of a solid without stretching is a perfectly well-defined thermodynamic transformation. Some concerns were expressed in the literature about the use of MD simulations to derive the relations between surface properties, in particular the analysis of the Shuttleworth Equation \cite{Gutman2016}. However, we will show that MD simulations and the statistical mechanics theory on which they are based, can give precise meaning to the surface properties and the relations between them.

\subsection{Statistical Mechanics description}

Before starting with the derivations of the definitions of SFE and SS in an atomistic system, we will give some definitions of the physical quantities which are needed in the rest of the Section. 

The total potential energy on an atomistic system is given by the sum of the interactions among atoms (for a more detailed exposition see \cite{Frenkel2001}). In this work we assume, for simplicity, that the system potential energy is defined in terms of a pair potential
\begin{equation}\label{Eq:pairPot}
\pot^{TOT}(\bfr) = \sum_l\sum_{n>l} U(r_{ln}),
\end{equation}
where $\bfr = \{\bfr_n\}$ represent coordinates of all the particles and $r_{ln} = |\bfr_l - \bfr_n|$ is the distance between particles $l$ and $n$. 

The stress tensor in a sample can be also defined in terms of relevant atomistic quantities. When a deformation occurs, the atoms change their relative position within the material, resulting in an imbalance of forces acting between the atoms. The fact that the state of stress, in general, depends on the directions stems from the fact that the deformation has directionality. The macroscopic state of stress of a material was put in relation with the atomistic structure of the matter by \citet{Clausius1870} through the Virial Theorem which we briefly introduce here.  It can be shown that the pressure tensor $\Ptens$ is equal to \citep{Evans1979,Thompson2009,Hummer1998}:
\begin{equation}\label{Eq:VirTheo}
	\Ptens V = \sum_{l} m_l\bfv_l\otimes \bfv_l + \frac{1}{2}\sum_{l} \bfr_{l} \otimes \bfF_l
\end{equation}
where $m_l$, $\bfv_l=\{\vx_{l}, \vy_{l}, \vz_{l}\}$, $\bfr_l=\{\x_{l}, \y_{l}, \z_{l}\}$, $\bfF_l = \{F^1_{l}, F^2_{l}, F^3_{l}\}$  are the mass, velocity, position and total force for atom $l$, and the double subscript indicates the $i$-th component of the $l$-th atom. The sum runs over all the atoms in the systems. In this work, all the sums have to be intended as running over the total number of atoms in the system.
For a pair potential, $\pot^{TOT}(\{\bfr_{l}\}) = \sum_l\sum_{n>l}U(r_{ln})$, where $r_{ln}= |\bfr_l - \bfr_n|$ and $F^i_{l} = -\totd{}{x^i_{l}}\pot^{TOT} = \sum_{ n } \cip{\frac{x^i_{ln}}{r_{ln}} \Dpot\cip{r_{ln}}} = \sum_{ n } \cip{x^i_{ln} F^i_{ln}}$, where we introduced $\Dpot(r)$ denoting the derivative of the function $U(r)$ with respect to its (scalar) argument, $F^i_{ln}$ as the $i$-th component of the force between atom $l$ and atom $n$ and $x^{i}_{ln}=x^i_{l}-x^j_{n}$ as the $i$-th components of the difference  $\bfr_l - \bfr_n$.

We can write for each component  $\p_{ij}$ of the pressure tensor $\Ptens$ in  \citep{Thompson2009}:
\begin{equation}\label{Eq:VirTheoComp}
        \p_{ij}V  =\sum_{l} m_l v^i_{l}v^j_{l} - \sum_{l}\sum_{n > l} \cip{\frac{x^{i}_{ln}x^{j}_{ln}}{r_{ln}} \Dpot\cip{r_{ln}}}.
\end{equation}
One consequence of the definition in \cref{Eq:VirTheo} that will be used in the following is that at zero temperature ($T=0$) the velocity of the atoms is zero, and the pressure tensor contains only the force term. 

In the following sections we will define the SFE and the SS via statistical mechanics quantity. Rigorously, these two objects are different from the ones defined in previous sections (i.e. \cref{Eq:sfe,Eq:SSsolid}), since they were defined for a macroscopic system using thermodynamics quantities. The goal of the analysis in the next sections is to show that indeed the SFE and SS defined using thermodynamics and statistical mechanics represents the same quantity and that they are related by an equation which we can identify with the Shuttleworth Equation. For this reason, we will use the same symbol $\gamma$ and $f^S$ to indicate them in the following derivation, by keeping in mind that, until the connection is established, they represents only the statistical mechanics version of the SFE and SS. 

\subsection{Zero Temperature}\label{Sec:zeroT}

\noindent For $T=0$, the entropic contribution in the definition of the Free Energies is zero. It then follows that the Helmolthz Free Energy $\HFE$ is equal to the internal energy of the system, which is, in turn, the sum of the interactions of the atoms in the simulation volume. Therefore for $T=0$ we can derive all the relevant equations for the ``surface'' properties by using only the potential energy.

In order to describe the SS and SFE, we need to calculate the potential energy in two geometries: the bulk, which we will denote with a subscript $b$, and the slab, denoted with a subscript $s$.

When considering potential energy (as in \cref{Eq:pairPot}), the material points can be associated with the particle positions and so the total potential energy of a deformed solid can be represented as $\pot^{TOT}(\hbfr)\equiv \pot^{TOT}(\bfr;\pbfu)\equiv \pot^{TOT,\e}$.   In what follows, to simplify the notations, we will indicate the strain state by including the superscript $u$  instead of the hat symbol (see discussion at the end of \cref{Sec:intro}). Then, when referring to the unstrained state, we will use the superscript $0$, as in $\pot^{TOT,0}$.

Without the loss of generality we will focus the discussion on the strain only in the $x$ direction, $\pbfu=(\e,0,0)$, unless otherwise specified, with the other cases being easily obtained by changing the relevant indexes. The SFE at $T=0$ is simply determined as the difference per unit area between the slab and bulk geometries:
\begin{equation}\label{Eq:T0gamma0}
	\gamma^0 \equiv \gamma(\pbfu=0) = \frac{\pot^{TOT,0}_s-\pot^{TOT,0}_b}{A^0}
\end{equation}
where $A^0$ is the area of the interface in the unstrained state, $\pot^{TOT,0}_s$ is the internal energy of the system in the slab configuration, $\pot^{TOT,0}_b$ is the internal energy of the system in the bulk configuration. We use the same symbol $\pot^{TOT}$ to represent the thermodynamic internal energy and the total potential energy of the system, since they are related, up to a constant, within the statistical-mechanical description. In the literature, it is usually assumed that $\gamma$ characterises the SFE of an unstrained solid and the superscript 0 is not included.  In general, however, $\gamma$ will depend on the strain state of the solid, as the result of the strain dependence of $\pot^{TOT,\e}$. We will write then:
\begin{equation}\label{Eq:T0gammau}
	 \gamma^\e \equiv \gamma(\pbfu) = \frac{\pot^{TOT,\e}_s-\pot^{TOT,\e}_b}{A^\e},
\end{equation}
with $A^u=A^0(1+\e)$ being the area of the strained surface, whose area is $A^0$ in the unstrained state. This definition of the SFE as a general function of the strain state of the system will be essential in the description of the SS. We will show that the relation in \cref{Eq:T0gammau} is consistent with our calculations where the SFE for different states of strain of the material are presented (see \cref{Sec:Results}).
 
The definition of the SS, $\surfst$, requires additional intermediate steps.  Being a tensorial quantity, its components need to be defined with respect specified coordinate axes. As stated before, we will consider only the $x$ direction of the SS tensor, $f_{11} \equiv f$, with a straightforward extension to the other directions. \\
We start by considering the work needed to stretch an already existing surface by a strain $u$, which is given by $u A^0 f$ and we derive its representation using microscopic quantities. \\
At zero temperature, we can write $\pot^{TOT,\e}_s - \pot^{TOT,0}_s$ as the work needed to stretch a slab (which represents a system with an interface) by a strain $\e$.  The stretching of the slab implies the stretching of the surface as well as of the volume. In order to isolate the work needed for the stretching of the surface, we can use the stretch of an equal volume of bulk and rewrite the quantity $\e A^0 f$ as an excess quantity between the slab and the bulk \citep{Muller2013}, similar to what we did for the SFE:
\begin{align}\label{Eq:fA0}
	\e A^0 f & = (\pot^{TOT,\e}_s - \pot^{TOT,0}_s) - (\pot^{TOT,\e}_b - \pot^{TOT,0}_b) \nonumber \\
	& = -(\pot^{TOT,0}_s-\pot^{TOT,0}_b) + (\pot^{TOT,\e}_s - \pot^{TOT,\e}_b)	 \nonumber \\
	& = - A^0 \gamma^0 + A^\e \gamma^\e.
\end{align}
From the above equation we see that the surface stress is a function of the strain $u$, i.e. $f(u)=\frac{1}{\e A^0}(- A^0 \gamma^0 + A^\e \gamma^\e)$. However, this expression cannot be directly used to calculate the surface stress at $\e=0$ which is the quantity we usually need. We will show at the end of our analysis how this problem can be circumvented (see \cref{Eq:FinUTI}). In what follows, we simplify the notation by not writing the dependence of $f$ on $\e$ explicitly, unless necessary.  In \citep{Bottomley2001}, the dependence of $f$ on $\gamma^\e$ alone was used to argue the validity of the Shuttleworth equation, which we don't believe is justified.

Now, we expand $\gamma^\e$ in Taylor series around the value $\e=0$:
\begin{equation}
	\gamma^\e = \gamma^0 + \cip{\totd{\gamma^\e}{\e}}_{\e=0} u + \lanO(\e^2)
\end{equation}
and we can neglect the terms $\lanO(u^2)$ thanks to the assumption of small strains (see discussion at the end of \cref{Sec:intro}). The SFE as a function of the strain and its Taylor expansion (up to the second order) was already considered by different authors \citep{Wolf1993,Fletcher2014}. For the purpose of this work, the Taylor expansion up to the first order will be enough. Thus we can write for $f$ (see Sec. S.3.L):
\begin{equation}\label{Eq:Shutt}
	f = \gamma^0 +  \cip{\totd{\gamma^u}{u}}_{u=0}.
\end{equation}
The same considerations can be done for the strain along the $y$ direction, i.e. $\e_{22}$ and for off-diagonal strains (i.e. $\e_{12}=\e_{21}$, see Section S.3.J of the SM).  
\Cref{Eq:Shutt} looks identical to the well-known Shuttleworth, however, we need to highlight here that we did not show yet that the quantities appearing here can be identified with the thermodynamic ones.

Let us now represent \cref{Eq:Shutt} in terms of the interactions between particles in the solid phase.  Again we consider the case $\pbfu=(\e,0,0)$.  We have (for a full derivation see Equation S.6 of the SM):
\begin{align}\label{Eq:dergamma}
	\totd{\gamma^\e}{\e} & = -\frac{\gamma^u}{(1+\e)} + \frac{1}{A^\e}\pard{}{\e} \cip{\pot^{TOT,\e}_s - \pot^{TOT,\e}_b} 
\end{align}
The final value of this derivative at $\e=0$ is (see Section S.3.E of the SM):
\begin{align}\label{Eq:derstr4}
	\cip{\totd{\gamma^\e}{\e}}_{\e=0} & = -\gamma^0 +  \frac{1}{A^0}\sum_{l}\sum_{n > l}\sqp{\cip{x_{ln}\force^1_{ln}}_{b} - \cip{x_{ln}\force^1_{ln}}_{s}}
\end{align}
If we compare \cref{Eq:Shutt,Eq:derstr4}, we see that the microscopic definition of the strain component $f_{11}$ is given by the last term in \cref{Eq:derstr4}, which is consistent with the virial expression for the excess pressure for $T=0$ (see \cref{Eq:VirTheo}). \Cref{Eq:derstr4,Eq:Shutt} represents the connection between the macroscopic thermodynamics description and the microscopic statistical mechanics description.

\subsection{Finite Temperature}

In previous Section we obtained the relation sought between the thermodynamics and the statistical mechanics of the interfaces. However, previous results was obtained under the rather restrictive hypothesis of $T=0$. In this Section we now generalize our discussion to finite temperature, i.e. $T > 0$. We proceed in the same way as \cref{Sec:zeroT}, by defining $\gamma^u$ and then deriving the components of the statistical mechanics version of the Shuttleworth equation $\gamma^0$, $f$, and $d\gamma/d\e$ at $\e=0$ for the system interacting via a given pair potential $\pot^{TOT}$. As a result of this section, we will show that the Shuttleworth equation is consistent if we calculate $\gamma^0$ from the thermodynamic path (bulk $\to$ slab) and $f$ from the integral of the stress profile across the interface, and that the statistical mechanics version of the Shuttleworth equation is equivalent with the known thermodynamic version, which therefore is proved.

At finite temperature, $\gamma^u$ represents the free energy difference per unit area between slab and bulk:
\begin{align}\label{Eq:SFEstr}
	\gamma^\e & = \frac{\HFE^u_s-\HFE^\e_b}{A^\e} 
\end{align}
where we used again the superscript $u$ to indicate that we are considering the strained configuration $\e$ (which again we assume  $\pbfu=(\e,0,0)$).
The SFE for the unstrained case can be easily obtained from \cref{Eq:SFEstr} by taking $\e=0$.

Following the same reasoning that brought us to \cref{Eq:fA0}, the SS at finite temperature is defined as:
\begin{align}\label{Eq:fFiniteT}
	\e A^0 f = (\HFE^u_s-\HFE^\e_b)-(\HFE^0_s-\HFE^0_b)  =  A^\e\gamma^\e - A^0\gamma^0 
\end{align}
from which we can obtain an equation similar to \cref{Eq:Shutt} but also valid for finite temperature. As we anticipated at the beginning of this Section, \cref{Eq:fFiniteT} represents the bridge between the thermodynamic and statistical mechanics. In fact, $(\HFE^\e_s-\HFE^\e_b)-(\HFE^0_s-\HFE^0_b)$ represents exactly the excess of Helmohltz Free Energy, $\HFE^\sigma_{exc}$, defined in \cref{Eq:fFreeE} and from \cref{Eq:fFiniteT} we will show that $f$ represents exactly the SS we defined through a thermodynamic route in \cref{Sec:LiqV,Sec:SolInt}.

In deriving $f$ from \cref{Eq:fFiniteT}, it turns out that there is a simpler thermodynamic path to follow. This new thermodynamic path still represents the difference between the SFE at strain $u$ and SFE in the unstrained configuration but  avoids the calculation of the SFE (i.e. it does not need to know the quantity $\gamma^u$). This is obtained by simply rearranging the terms in the second part of \cref{Eq:fFiniteT}:
\begin{align} \label{Eq:defShuttA}
	\e A^0 f  = A^\e\gamma^\e -A^0\gamma^0 =  (\HFE^\e_s-\HFE^0_s)-(\HFE^\e_b-\HFE^0_b)  
\end{align}
In this new path we look for the difference in Helmholtz Free Energy to strain a system (in bulk or slab configuration) from the unstrained state ($\e=0$) to strain equal to $\e$ :
\begin{align}\label{Eq:defNewTI}
	\Delta \HFE_{\substack{\mbox{slab}\\ \mbox{unstrained}}\:\:\rightarrow \:\:\substack{\mbox{slab}\\ \mbox{strained}}  }  = \HFE^\e_s-\HFE^0_s = 
			 \int_{0}^{\e} \de \nu\aver{\pard{\pot_s^{TOT,\nu}}{\nu}}_{\nu}  
\end{align}
where $\aver{\cdot}_\nu$ represents the canonical ensemble average of the system under the strain $\nu$, and analogously for the bulk configuration by replacing subscript $s$ with $b$. The last quantity in \cref{Eq:defNewTI} is the standard results for the difference in free energies between two state of the system as obtained within the framework of the Thermodynamic Integration theory, see e.g. \cite{Frenkel2001}.

We can therefore write:
\begin{align} \label{Eq:NewTIf}
	\e A^0 f  =  (\HFE^\e_s-\HFE^0_s)-(\HFE^\e_b-\HFE^0_b)=   \int_{0}^{\e} \de \nu\aver{\pard{}{\nu}\cip{\pot_s^{TOT,\nu}-\pot_b^{TOT,\nu}}}_{\nu}  
\end{align}
from which we can derive the expression of the SS as function of $\e$, $f(\e)$ (See Section S.3.H of the SM):
\begin{align}  \label{Eq:NewTIf2}
	 f(\e) & =\frac{1}{u A^0}  \int_{0}^{u} \frac{\de \nu}{(1+\nu)}\aver{\cip{\frac{1}{2}N\kb \mathcal{T}+ \sum_{l} \sum_{n >l}\cip{\hat{x}_{ln}\hat{\force}^1_{ln}(\nu)}_{b}}-\cip{\frac{1}{2}N\kb \mathcal{T}+ \sum_{l} \sum_{n >l}\cip{\hat{x}_{ln}\hat{\force}^1_{ln}(\nu)}_{s}}}_{\nu}  
\end{align}
where $\hat{\force}^1_{ln}(\nu)$ is the $x$-component of the force in the strained system identified by $\nu$, and $x_{ln}=\hat{x}_{ln}/(1+\nu)$ where $\hat{x}_{ln}$ represents the difference in the $x$ coordinate between atom $l$ and atom $n$ in the strained configuration (for a discussion and a definition of coordinates in a strained configuration see \cref{Eq:HatDef}).

The SS usually considered in calculations is the SS in the unstrained configuration (i.e. $\e=0$). However, the previous expression for $\e=0$ gives the indeterminate quantity $0/0$. To derive the value of SS at $\e=0$ we proceed in the following way. We use the Fundamental Theorem of Calculus to rewrite \cref{Eq:NewTIf2} as a differential relation between the SS, the SFE (see \cref{Eq:defShuttA}) and the argument of the integral:
\begin{align}
  	\totd{(\e A^0 f(\e))}{\e} = \totd{A^u \gamma^\e}{\e}= \frac{1}{(1+u)}\aver{\cip{\frac{1}{2}N\kb \mathcal{T}+ \sum_{l} \sum_{n >l}\cip{\hat{x}_{ln}\hat{\force}^1_{ln}(u)}_{b}}-\cip{\frac{1}{2}N\kb \mathcal{T}+ \sum_{l} \sum_{n >l}\cip{\hat{x}_{ln}\hat{\force}^1_{ln}(u)}_{s}}}_{u}  
\end{align}
If we calculate the quantities in the previous equation at $u=0$ we obtain the value of SS we needed:
\begin{align}\label{Eq:FinUTI}
 	 f\equiv f(0)  =  \frac{1}{A^0}  \left.\frac{d A^\e \gamma^\e}{d \e}\right|_{\e=0} & =  \frac{1}{A^0} \aver{\cip{\frac{1}{2}N\kb \mathcal{T}+ \sum_{l} \sum_{n >l}\cip{x_{ln}\force_{ln}}_{b}}-\cip{\frac{1}{2}N\kb \mathcal{T}+ \sum_{l} \sum_{n >l}\cip{x_{ln}\force_{ln}}_{s}}}_{u=0} \nonumber \\
 	 &  =  \aver{\int_{-\infty}^{+\infty}{(\p^\alpha_{11} - \p^\sigma_{11}(x^3))\de x^3}  }_{\e=0} 
\end{align}
The relation just derived represents the link between the thermodyamics and the statistical mechanics definition of the quantity $f$. This link is represented by the the microscopic definition of the stress within a sample and their identification with the macroscopic stresses (see \cref{Eq:VirTheo,Eq:VirTheoComp}).

With \cref{Eq:FinUTI}, we showed that \cref{Eq:Shutt} does represent the Shuttleworth equation, which we derived following a statistical mechanics route, instead of the thermodynamics one.

\section{Computational model}\label{Sec:Compcleav}

In this section we will describe how the different quantities can be calculated in a MD simulations.

\subsection{Cleaving} \label{Sec:cleav}

The concept of ``cleaving'' was introduced by \citet{Eriksson1969} and used in several other works \cite{Haiss2001,Cammarata1994}. Its meaning was disputed in the literature \cite{Marichev2009,Marichev2010,Marichev2009b,Marichev2011} on the basis that no clear definition of the reversible cleaving process was given. However, it can be rigorously defined within the framework of the statistical mechanics theory and actually computed in MD simulations. 

Reversible cleaving in molecular dynamics dates back to \citet{Miyazaki1976} and was later used by \citet{Broughton1986} who presented a thermodynamic transformation to reversibly create an interface in a Lennard-Jones face-centered-cubic (fcc) crystal-liquid system. The methodology was further extended in \citep{Davidchack1998,Davidchack2000,Davidchack2003} and we will give here a brief description of it. 

For the calculation of the excess free energy of the interface between two phases $\alpha$ and $\beta$ (which could be the same thermodynamic phase, e.g. liquid, of two different materials), the most general formulation of the cleaving methods includes four steps:
\begin{itemize}
	\item \textit{step1}: The cleaving potential is introduced in the phase $\alpha$ creating a precursor to the interface with $\beta$
	\item \textit{step2}: The cleaving potential is introduced in the phase $\beta$ creating a precursor to the interface with $\alpha$
	\item \textit{step3}: The boundary conditions are rearranged and the interactions are switched on between the phases
	\item \textit{step4}: The cleaving potentials are gradually removed from the new system with phase $\alpha$ and $\beta$ in contact
\end{itemize}

\subsubsection{Simplified Thermodynamic Path}\label{Sec:wells}

For the calculation needed in this work we use a simplified thermodynamic path derived from the one shown in the previous section. Since we have only one phase (the crystal phase, $\alpha$) we can consider three steps only:
\begin{itemize}
	\item \textit{step1} (S1): The cleaving potential is introduced in the bulk of the phase $\alpha$, creating a precursor to the interface
	\item \textit{step2} (S2): The interactions between the two sides of the crystal at the interface are switched off
	\item \textit{step3} (S3): The cleaving potentials are gradually removed from the slab of the $\alpha$ phase
\end{itemize}
The cleaving potential chosen for this calculation is represented by the wells model \cite{Handel2008} which is conceptually similar to the walls model\citep{Davidchack2003}. In both cases we put on each side of the cleaving cut two planes with interaction sites that interact with atoms through a particular potential. The main difference with the walls method is that the two planes are now fixed and the strength of the well potential, $\psi(r;\lambda)$, is the quantity which is varied within the simulation. 
The well potential we use in this work is  defined as \cite{Handel2008}:
\begin{equation}\label{Eq:wells}
	\psi(r;\lambda) = 
		\begin{cases}
			\lambda d_w\sqp{\cip{\frac{r}{r_w}}^2-1}^3 \:\:\: r < r_w \\
			0  \:\:\: r \geq r_w
		\end{cases} 
\end{equation}
where $r=\abs{\bfr-\bfR}$ is the distance between the atom in position $\bfr$ and the centre of the well in position $\bfR$, $d_w$ is the well depth, $r_w$ is the cut off radius and $\lambda\in[0,1]$ is the parameter that varies from zero to one as the wells potential is switched on in \textit{step1}. In the wells version each atom interacts with \textit{both} sides of the cleaving plane, therefore the total interaction felt by each atom is just the sum of all the interaction points on both sides of the cleaving plane:
\begin{equation}
	\Psi(\bfr;\lambda) = \sum_{j}{\psi\cip{\abs{\bfr-\bfR_j}; \lambda}}.
\end{equation}
The wells are located on the ideal crystal lattice positions on the first and last crystal layer of the box in direction $z$.

In \textit{step2}, we gradually switch off the interaction between the two sides of the crystal. We modify the size of the box in the direction $z$ by adding a quantity $z^F_w$ to it. In this way we are increasing the distance over which the atoms on the two sides of the boxes are interacting with their periodic images. Finally, the cleaving potential is removed in \textit{step3} by varying $\lambda$ from one to zero.

The SFE is given by the sum of the reversible work, $w$, done in each of the three steps. In S1 and S3 the reversible work is calculated by integrating over the parameter $\lambda$:
\begin{equation}
		w_{S1} = \int_{0}^{1}{\aver{\pard{\Psi(\bfr;\lambda)}{\lambda}} \de \lambda}\,,\;\; w_{S3} = -\int_{0}^{1}{\aver{\pard{\Psi(\bfr;\lambda)}{\lambda}} \de \lambda}
\end{equation}
The reversible work in S2 is obtained from the total potential of the system, $\pot^{TOT}(r; z_w)$,
\begin{equation}\label{Eq:s2} 
	w_{S2} = \int_{0}^{z_w^{F}}{\aver{\pard{\pot^{TOT}(r; z_w)}{z_w}} \de z_w}
\end{equation}
The total potential of the system now includes $\lambda$ and $z_w$ as parameters, the former through the wells potential:
\begin{equation}
		\pot^{TOT}(r; \lambda, z_w ) =   \sum_{i=1}^{N}\sum_{ j > i}U\cip{r_{ij};z_w}  + \sum_{k=1}^{M}\Psi(\bfR;1)
\end{equation}
with $N$ number of atoms and $M$ number of wells respectively.
It can be shown that (see Section S.3.G of the SM):
\begin{align}
	\pard{}{z_w} U\cip{r_{ln};z_w} & = \force^3_{ln}\cip{z_w}
\end{align}	
where $\force^3_{ln}$ is the component of the force between atoms $l$ and $n$ in the $z$-direction. The integral in \cref{Eq:s2} becomes:
\begin{equation}
	w_{S2} = \int_{0}^{z_w^{F}}{\aver{\sum_{\substack l \\ n > l }{}^\prime\force^3_{ln}\cip{z_w}} \de z_w}
\end{equation}
where the symbol $\sum^\prime_{\substack l \\ n > l }{}$ is used to indicate that the sum of the forces must be restricted to those pair interactions crossing the interface. 
The simplified path for cleaving described in this section also shows clearly why the Shuttleworth equation contains two terms, one related to the creation of the interface ($\gamma$) which applies to liquids and solids, and the second term (the derivative of $\gamma$ with respect the strain) which applies only to solids. 
The cleaving considers only the interactions between two adjacent planes that need to be removed in order to create a new surface, i.e. the cleaving model allows to calculate $\gamma$ only. However, when the system is cleaved at finite temperature, it is allowed to relax at each step. During this relaxation the configuration of the atoms on the surface changes, the way this change happens is different for liquids and solids and this is the reason behind the particular behaviour of the interface properties in solids and liquids.

\subsection{Surface Stress for liquids}

In MD simulations the component of the tensor stress are calculated by using \cref{Eq:VirTheo}. The surface stress is then usually calculated by using the following equation:
\begin{equation}\label{eq:statistics}
 f^S=\gamma=\int_{-\infty}^{+\infty}{ \de z\cip{P-\frac{1}{2}\cip{\p_{11}(z)+\p_{22}(z)}}} 
\end{equation}
which seems to be different from the one we reported for liquids (see \cref{Def:surfstLVap}). However, we need to remember that $\p_{11}(z) = \p_{22}(z) = \p(z)$ which return the same equation we reported. The reason to consider this alternative version of the equation is that in a MD simulations $\p_{11}(z)$ and $\p_{22}(z)$ are independent measurements. Using their average in \cref{eq:statistics} increases the precision of the results (i.e. we are calculating the surface stress by averaging over more observations).

\subsection{Computational Details}

\noindent All the simulations are performed in LAMMPS \citep{Plimpton1995,Plimpton2007}. In the following discussion we will use Lennard-Jones reduced units and we will assume $\kb=1$ for the Boltzmann constant. The time step used is $\Delta t=0.005$ \citep{Davidchack2003} and all the simulations were performed in NVT ensemble. The atoms interact with the Broughton-Gilmer modified potential \citep{Broughton1983} (see Equation S.11 in SM) which was implemented as a new pair style in LAMMPS.
We simulated a fcc structure in three orientations: (100), (110), (111) at four different temperatures: $T=0$, $0.1$, $0.2$, $0.3$.
We calculated $\gamma$ by using the simplified cleaving path described in \cref{Sec:cleav}. The stress tensor $\surfst$ was calculated with the system in the \textit{slab} configuration, with the surfaces aligned with the z-direction. 
Simulations in the slab configurations are equilibrated for 500000 time steps after which a production run of 2000000 time steps follows. Quantities are saved every 10000 time steps. 

The goal of the simulations is to verify the Shuttleworth equation and the expression for the surface stress obtained in previous sections.  In particular, we calculate the surface stress in 11 and 22 directions by using the expression shown in \cref{Eq:SSsolid} where $P$ is the bulk pressure and $\pi_{ii}(z)$ is the stress profile calculated during the simulation using the virial expression (see \cref{Eq:VirTheo}). In a slab simulation we have two surfaces and we consider as bulk pressure the average of the pressure tensor in the middle of the box at sufficient distance from the interface such that all the disturbances caused by the interface are smaller than the statistical uncertainty due to atomistic fluctuations.
We calculate all the terms of the Shuttleworth equation by considering, for each system at each temperature, different strain states of the crystal. The derivative of the SFE with respect the strain is then obtained by numerical calculation of the tangent in the point $\e_{ii}=0$ (see Section S.2.A of the SM for more details). 
The strained system is built by simply multiplying the relevant coordinate ($x$ or $y$)  of the unstrained system by $(1+\e_{ii})$ where $i=1,2$, for strains in the direction $x$ and $y$, respectively. We considered 51 strains equally spaced in the interval $[-0.005,0.005]$.

For each system we then proceed to calculate $\gamma^{\e_{ii}}$. This procedure is similar to the one reported in the literature by different authors  \citep{Wolf1993,Hecquet2018,Price1976}, who, however, consider only the case at zero temperature. More details for the calculations are reported in the Supplemental Material (SM, see Section S.2). 

\clearpage

\section{Results} \label{Sec:Results}

\noindent In this Section we present the results obtained using the framework developed in the previous parts of the paper. We start our discussion by considering the zero temperature systems and then provide results for finite temperature.

\subsection{Zero Temperature}

\begin{figure}[H]
		\includegraphics[width=\textwidth]{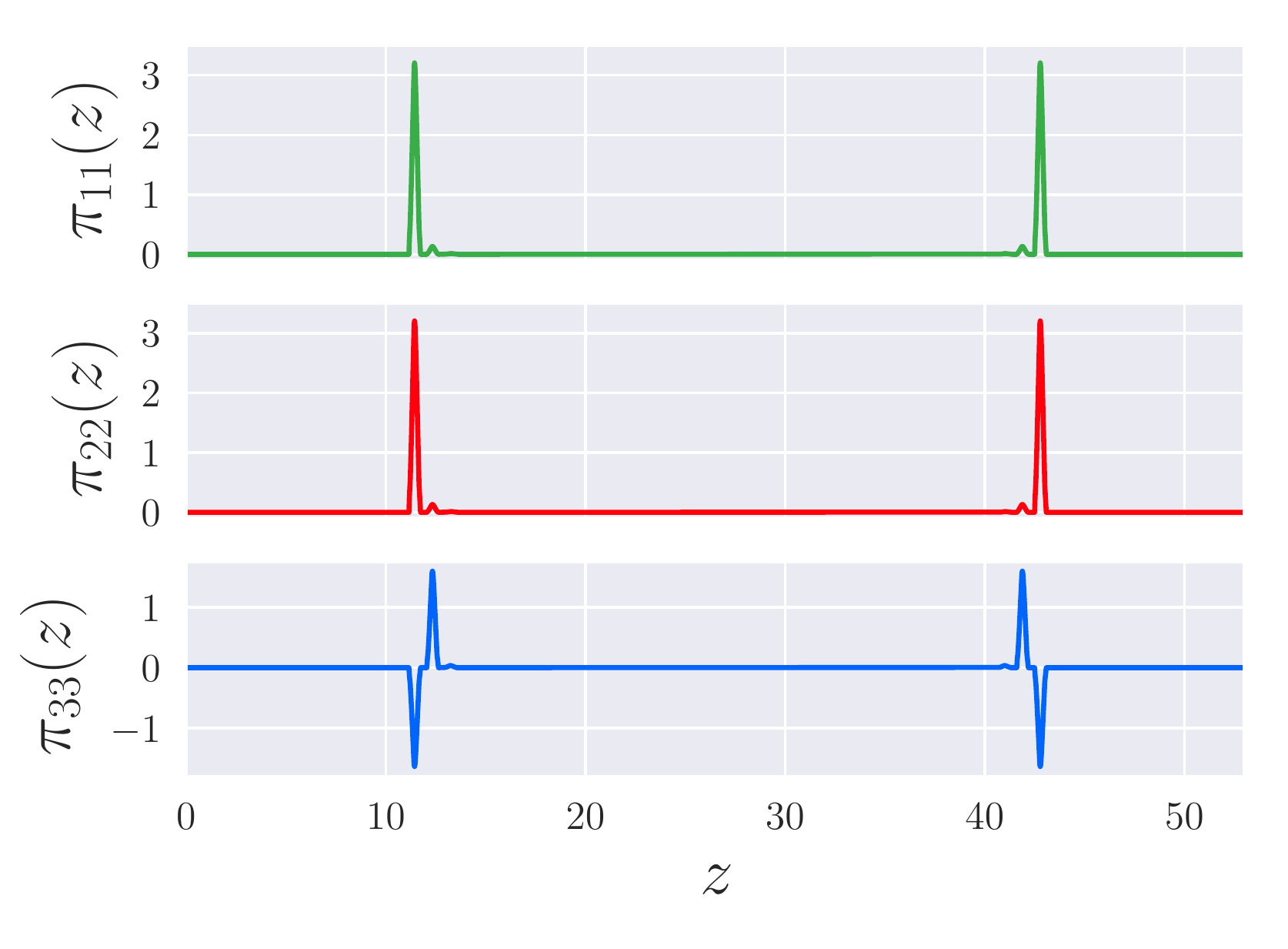}
		\caption{Pressure components for the fcc crystal in orientation (111) at $T=0$ in the three directions $x$, $y$ and $z$}
		\label{Fig:stress111T0}
	\end{figure}

\noindent In \cref{Fig:stress111T0} we reported the pressure profile along the three directions $x$, $y$, $z$, (i.e. the three components of the main diagonal of the pressure tensor, $\Ptens$) for the orientation (111). We first want to highlight a fact related to the $z$ components of the pressure tensor. In $x$ and $y$ directions we can clearly see an excess of the stress with respect the centre (i.e. the bulk) of the crystal sample which represents in both case a compressive stress (negative stress or positive pressure). In $z$ directions near the interface we observe a deviation from the bulk value, but it seems completely symmetric around the bulk pressure value. This last observation can be made more quantitative by calculating the integral of pressure profile in the pressure component $\p_{33}$ along $z$.  We obtain the value of the order of $10^{-6}$  which is close enough to zero to be ascribed to the finite precision of the calculation and be considered exactly zero. 
The presence of the stress in the $z$ direction stems from the rearrangement of the atoms in the surface layers with respect the bulk, which produces the surface excess quantities in $x$ and $y$ directions. However, the mechanical equilibrium assumption prevents the presence of any net stress on the $z$ direction, which therefore must sum to zero.
As anticipated in \cref{Sec:Liq-vap}, we obtain that the stress tensor in the $z$-direction cannot be considered an excess quantity and it should not be considered in the equation related to calculation of the interface properties. Or, in different terms, expressions for the interface quantities do not depend on any normal component with respect the surface. 
Again, we stress here that despite this different interpretation, the results usually considered using the ``mechanical definition'' are not affected because for liquids and hydrostatic solids there is no difference in the final equations between the bulk pressure and pressure along the normal to the interface ($\p_{33}$ in this case). However, we suggest here to avoid the definition of the surface properties in terms of normal component of the stress, which could eventually bring some confusion when the directions of the stress matter (i.e. for the case of a non-hydrostatically stressed solid). 
In \cref{Fig:stress111T0} the directions $x$ and $y$ show the excess quantity which must be considered for the calculation of the surface stress. 

\begin{figure}[H]
	\centering
			\includegraphics[scale=0.7]{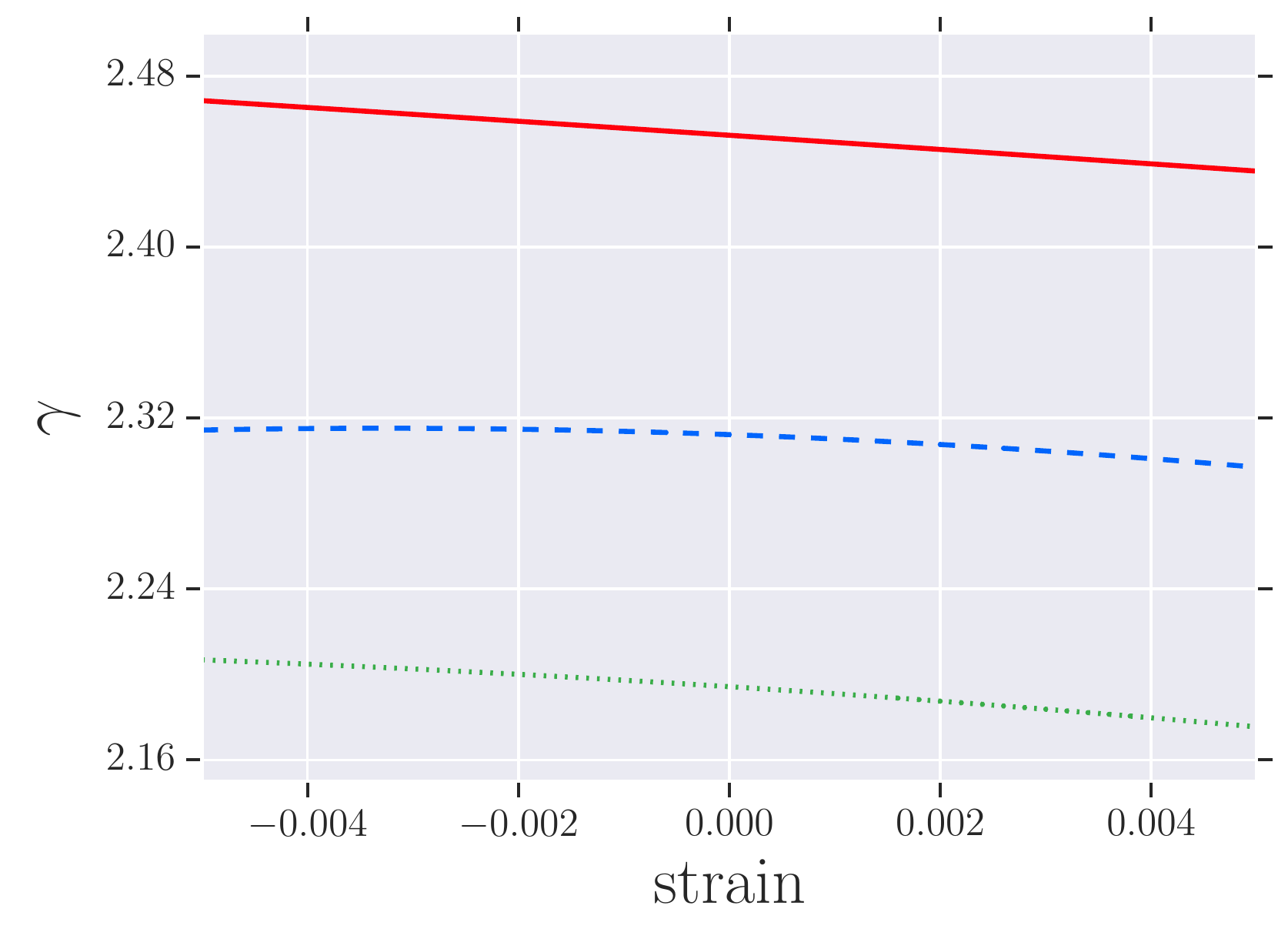}	
			\includegraphics[scale=0.7]{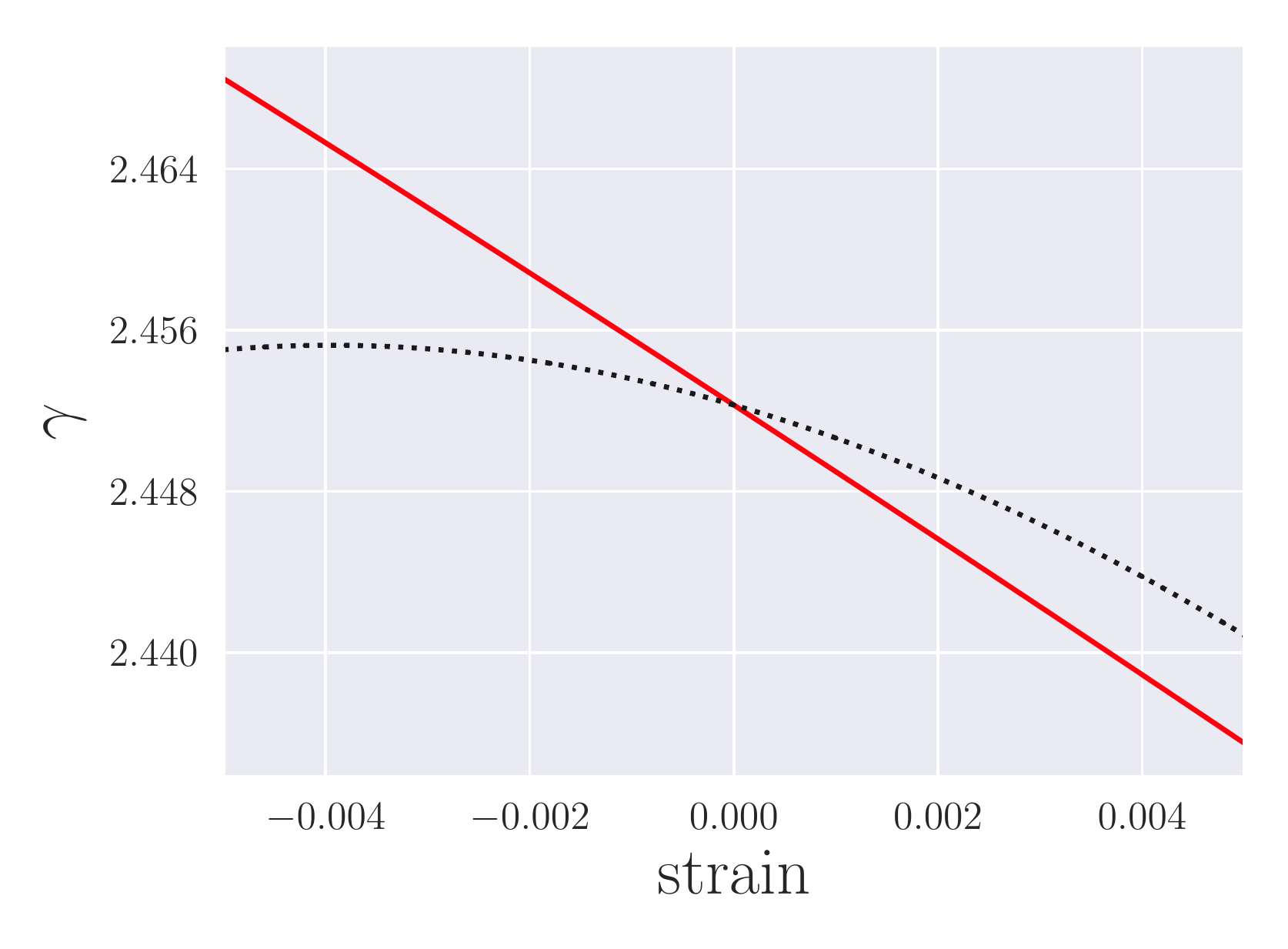}	%
		\caption{Top: Surface free energy for the fcc crystal in orientation (111) (dotted green line), (110) (solid red line) and (100) (dashed blue line) at $T=0$ as function of the strain in the $x$ direction. Bottom: Surface free energy for the fcc crystal in orientation (110) at $T=0$ as function of the strain in the $x$- (solid red line) and $y$-direction (dotted black line)}
		\label{Fig:T0gamma}
	\end{figure}

We then proceed to calculate the SFE $\gamma$ using the cleaving method described in \cref{Sec:cleav}. We calculated $\gamma$ as a function of the strain in the $x$ and $y$ directions and results for the three orientations are shown in \cref{Fig:T0gamma}.
The SFE for zero strain is reported in \cref{Tab:T0} for all the orientations considered. \citet{Broughton1986IV} reported the values of $2.1921, 2.4500,  2.3100$ for the SFE in the (111), (110) and (100) orientations at zero temperature, respectively. The small difference between our results and theirs can be attributed to the systematic errors due to size effects since the systems sizes in \citep{Broughton1986IV} are smaller than in our simulations. 
From the value of the SFE shown in \cref{Fig:T0gamma} (top picture) we can calculate (see Section S.2.A of the SM) the derivative with respect the strain, (i.e. the term $\cip{\pard{\gamma^u}{u}}_{u=0}$ in the Shuttleworth equation, see \cref{Eq:Shutt}). For orientations (100) and (111)  the SFE is the same for the strains in $x$ or $y$ directions. This implies that the value of the derivative will be the same for both directions, and we obtain the equality of the stress tensor in the $x$ and $y$ directions. The orientation (110) does not show this symmetry and straining the system in the two different directions ($x$ and $y$) gives different values of the SFE. These results are reported in \cref{Fig:T0gamma}  where we note, as expected, that the two curves intersect when $\e_{11}=\e_{22}=0$. It is clear that the value of the derivative is different in the two direction for (110) case. A summary of all the terms of the Shuttleworth equation is reported in  \cref{Tab:T0}. All results are consistent with the Shuttelworth equation, with small deviations stemming from the precision of the derivative estimation. 
\begin{table}[ht]
\begin{center}
  \begin{tabular}{  c | c | c | c | c  }
    \hline
     \multirow{ 2}{*}{Orientation }& $\gamma$ & $\pard{\gamma}{\e_{ii}}$ & $f_{ii}$ & $\gamma+\pard{\gamma}{\e_{ii}}$ \\ 
				& $(\epsilon \sigma^{-2})$ & $(\epsilon \sigma^{-2})$ & $(\epsilon \sigma^{-2})$& $(\epsilon \sigma^{-2})$ \\ \hline
     (111)      &  $2.194280$ & $-3.141741$  & $-0.947707$  &  $-0.947461$  \\  
     $(110)_{\e_{11}}$ &  $2.452284$ & $-3.300510$  & $-0.848423$  &  $-0.848226$ \\ 
     $(110)_{\e_{22}}$ &  $2.452284$ & $-1.467676$  &  $0.983892$  & 0.984608 \\ 
     (100)      & $2.312219$ & $-1.793525$  &  $0.516164$  & $0.518694$ \\                        
    \hline
  \end{tabular}
\end{center}
  \caption{Summary of the results of the different terms of the Shuttleworth equation obtained at zero temperature for all the directions and orientations considered. For (100) and (111) orientations, we report results with the stretching in $x$ direction only. For (110) we report the results in both the principal directions ($x$ and $y$, $i=1,2$)}
 \label{Tab:T0}
\end{table}
%
\newpage

\subsection{Finite Temperature}

\begin{figure}
\begin{tabular}{cc}
 \includegraphics[scale=0.8]{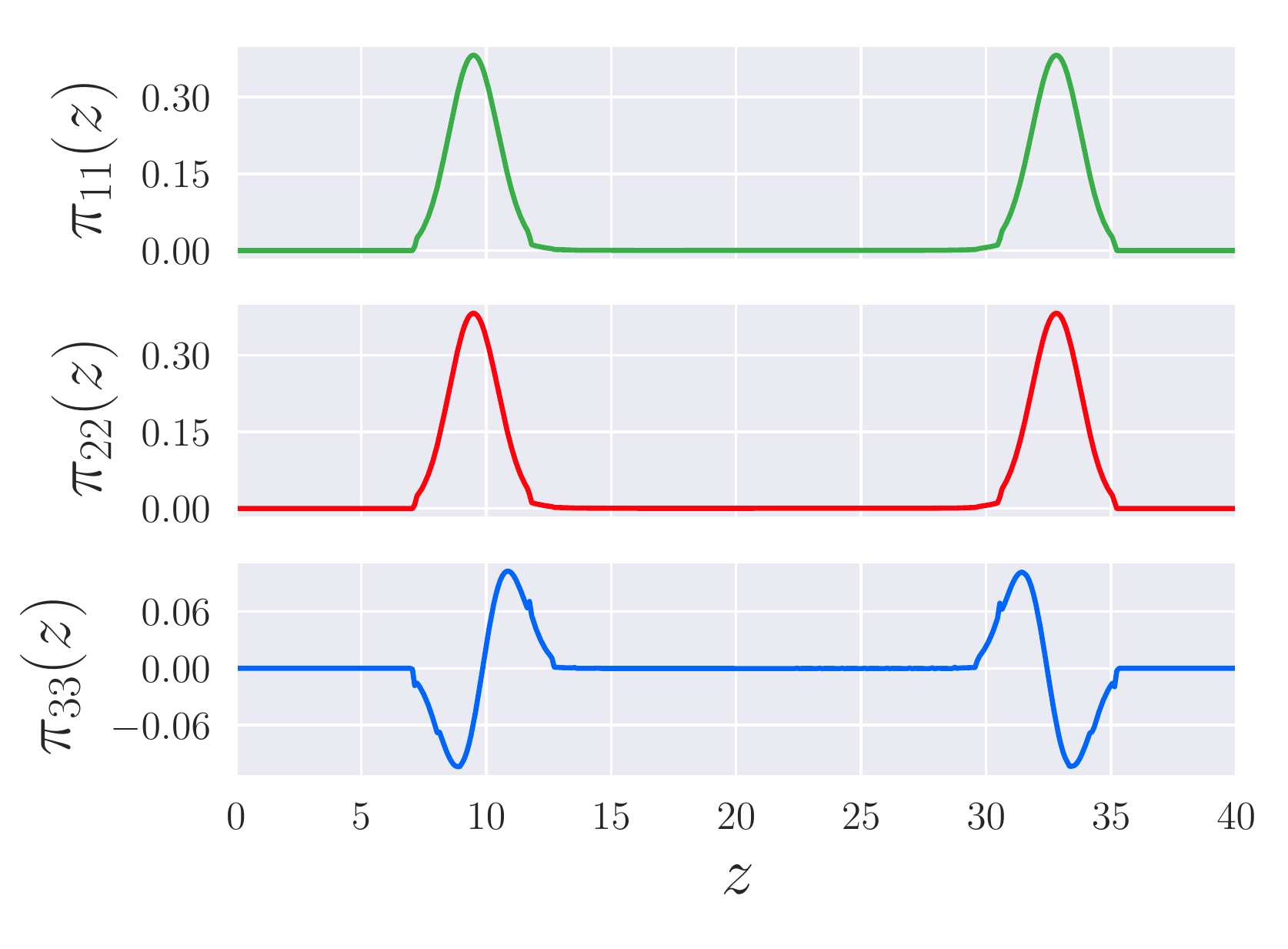} \\
 \includegraphics[scale=0.8]{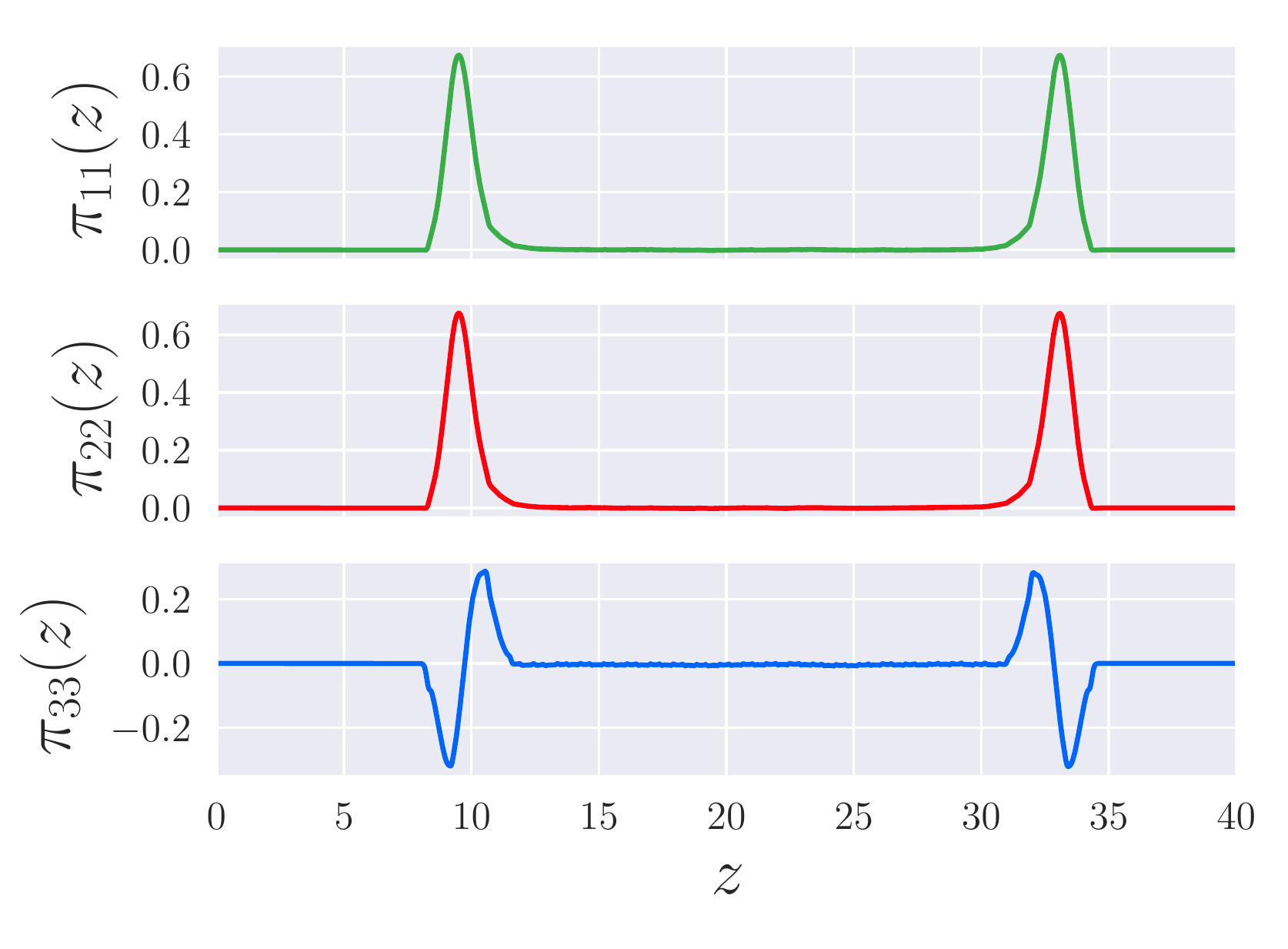} \\
\end{tabular}
\caption{Top: pressure profile for (111) at temperature T=0.1$\epsilon$ in the three directions. Bottom:  pressure profile for (111) at temperature T=0.3$\epsilon$ in the three directions.}
\label{Fig:PressCum}
\end{figure}

\noindent The pressure profiles for orientation (111), in the three different spatial directions, and temperatures $T=0.1\epsilon$ and $T=0.2\epsilon$ are shown in \cref{Fig:PressCum}.

From \cref{Eq:SFEstr} we can define the contribution of the internal energy $U$ and the entropy $S$ to the SFE by:
\begin{align}\label{Eq:entropy}
	\gamma^u & = \frac{\HFE^u_s-\HFE^u_b}{A^u} \nonumber \\
		     & = \frac{1}{A^u}\cip{U^u_s-U^u_b-TS^u_s+TS^u_s} \nonumber \\
		     & = \frac{1}{A^u}\cip{\Delta U^u - T\Delta S^u} =  \Delta \mathfrak{v}^u - T\Delta \mathfrak{s}^u
\end{align}
where the superscripts and subscripts have the usual meaning of system in bulk or slab configuration in the state of strain identified by $u$,  $\Delta U^u$ and $\Delta S^u$ describe the difference between bulk and slab of internal energy and entropy respectively for the strain state $u$ while $\Delta \mathfrak{v}^u$ and $\Delta \mathfrak{s}^u$ represent the same quantities per unit area. The internal energies $U^u_s$ and $U^u_b$ are calculated as the NVT ensemble averages of $\pot^{TOT,u}_s(r)$ and $\pot^{TOT,u}_b(r)$ respectively. In our systems there are always two interfaces, therefore the area $A^u$ is defined as $A^u=2L^u_xL^u_y$ where $L^y_x$ and $L^y_x$ are the size of the strained system in the $x$ and $y$ directions respectively. We use \cref{Eq:entropy} to estmate the value of the entropy $\Delta \mathfrak{s}^u$ at each temperature, knowing the SFE $\gamma$ and the internal energy $\Delta \mathfrak{v}$ at that temperature.

\begin{figure}[H]
	\centering
			\includegraphics[scale=0.5]{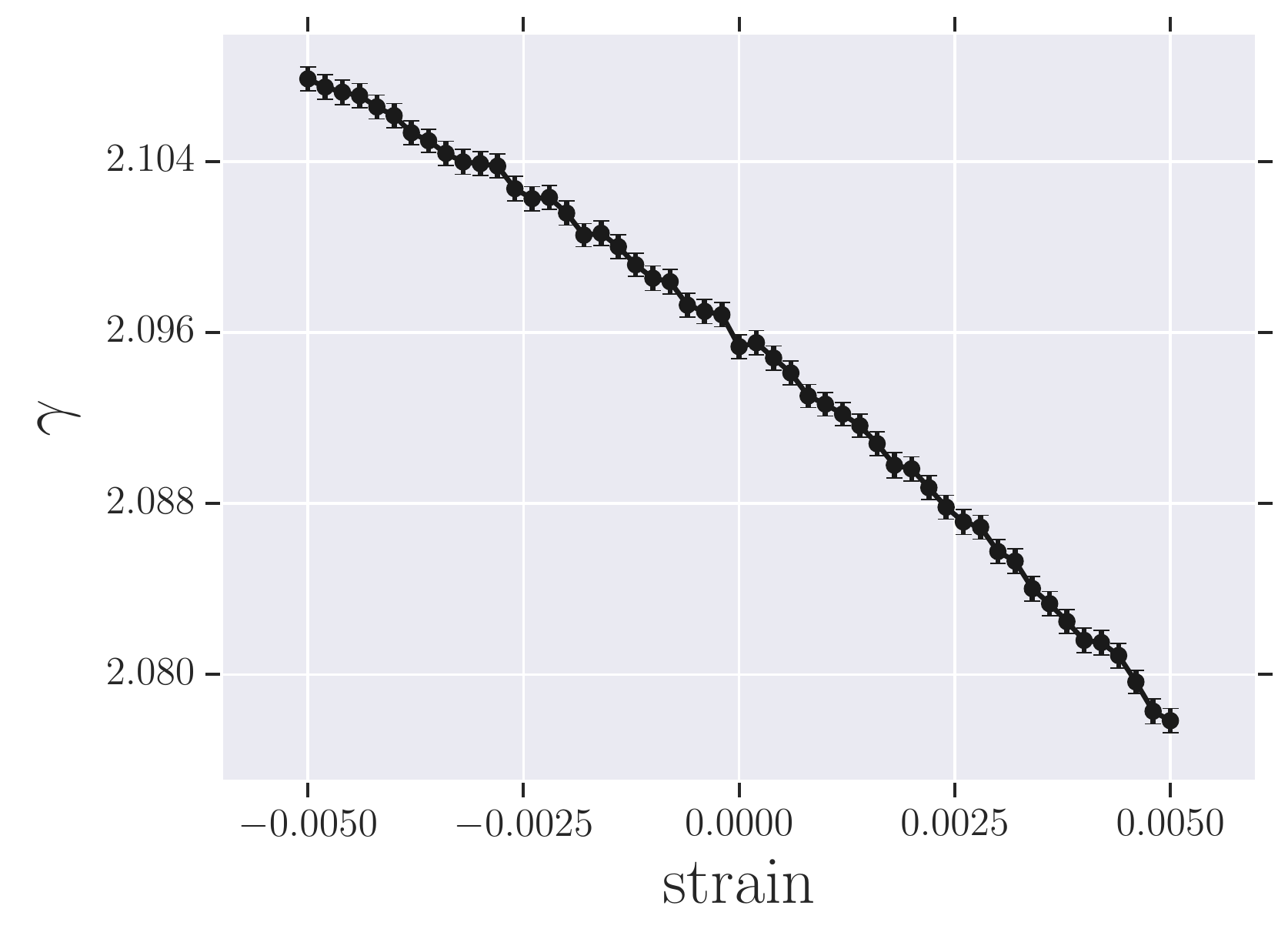}	
			\includegraphics[scale=0.5]{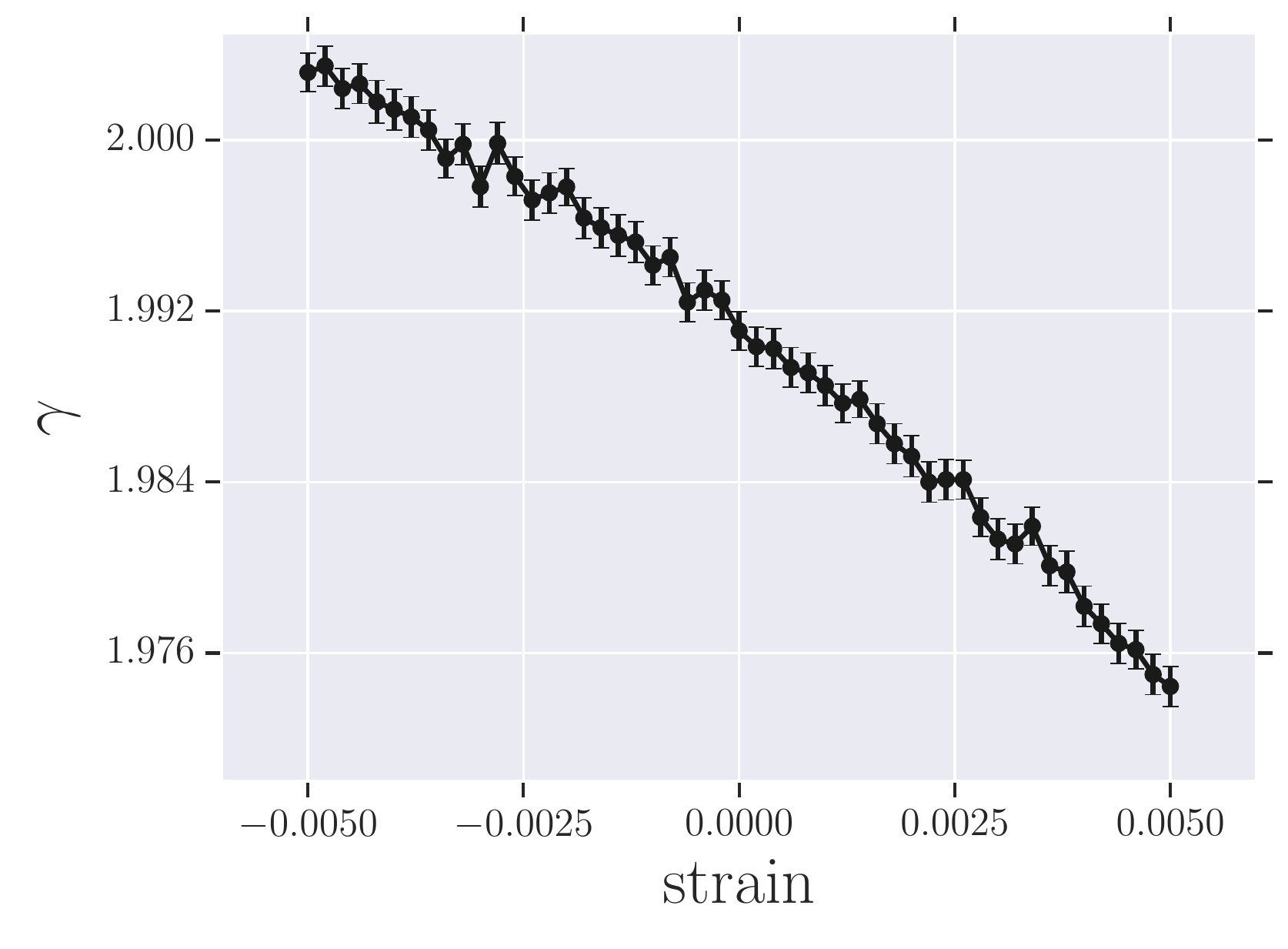}	
			\includegraphics[scale=0.5]{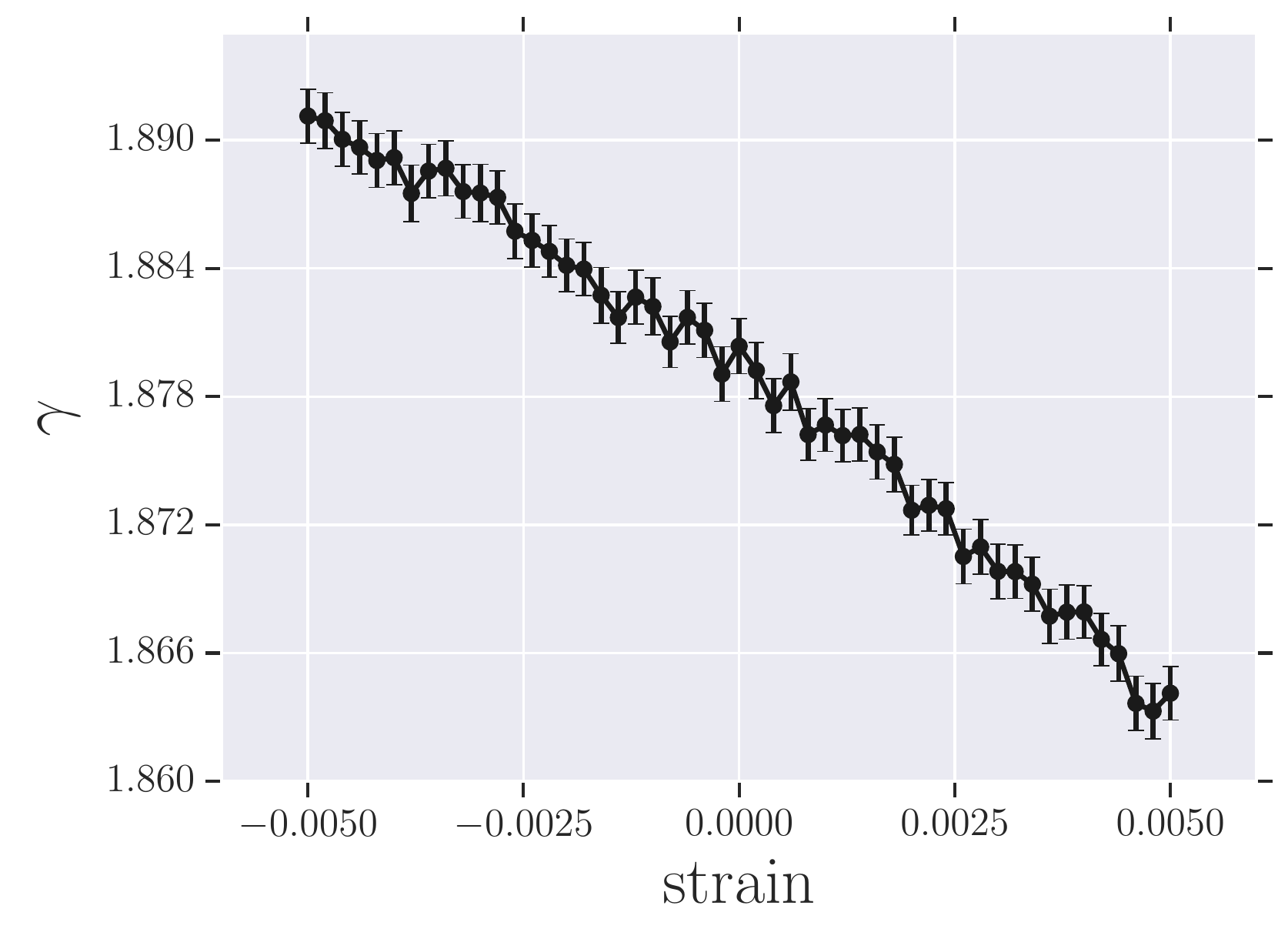}	
		\caption{From top to bottom, SFE $\gamma$ as a function of the strain rate in the $x$ direction for the orientation (111) at three temperatures, T= 0.1, 0.2, 0.3$\epsilon$. }
		\label{Fig:gammaFiniteT}
	\end{figure}

The effect of the increase of the temperature can be observed in the calculation of the SFE at different strain rates, which is shown in \cref{Fig:gammaFiniteT}. The error associated to the SFEs at each strain increases with the temperature, as expected from the greater mobility of the atoms at higher temperature.  However, while the error for the value of the SFE remains relatively low for the whole range of temperatures considered (smaller than 1$\%$, see also \cref{Tab:exx111}) the error associated with the estimation of the derivative component of the Shuttleworth equations can be an order of magnitude larger.  
At the highest temperature we considered (T=0.3$\epsilon$) other effects can influence the calculation of the SFE. Some atoms can leave the surface and the subsequent surface reconstruction can modify the value of the SFE, leading to possible systematic errors in the results. 
However, even for this temperature, which is well below the melting temperature of T=0.617$\epsilon$ \citep{Davidchack2003}, the vapour pressure can be still considered negligible. This can be observed in the pressure profile of a sample of crystal in the slab configuration, as seen in \cref{Fig:PressCum}.

A summary of the results for all the temperatures considered are reported in  \cref{Tab:exx111} for the orientation (111) direction. The other directions are reported in the SM (see Table S-I to Table S-VI).

\begin{minipage}{\linewidth}
\centering
\begin{center}
  \begin{tabular}{ l | c | c | c | c | c | c  }
      \toprule[1.25pt]
       \multicolumn{6}{c}{(111)}  \\
    \toprule[1.25pt]
     T & $\gamma$ & $\pard{\gamma}{\e_{11}}$ & $f_{11}$ & $\gamma+\pard{\gamma}{\e_{11}}$ & $ \Delta \mathfrak{v}^0$ & $ \Delta\mathfrak{s}^0$ \\
           ($\epsilon)$ & $(\epsilon \sigma^{-2})$ & $(\epsilon \sigma^{-2})$ & $ (\epsilon \sigma^{-2})$ & $(\epsilon \sigma^{-2})$ & $(\epsilon)$ &  \\ 
     \hline
     0    &  $2.194280$ & $-3.141741$  & $-0.947707$  &  $-0.947461$  & 2.194280 &  - \\  
     0.1 & $2.0905(6)$ & $-3.01(3)$ & $-0.9235(4)$  &  $-0.91(3)$ &  $2.1635(4)$ & $0.730(7)$\\ 
     0.2 & $1.9911(9)$ & $-2.88(5)$ & $-0.913(5)$ & $-0.89(5)$ & $2.138(1)$ & $0.737(7)$\\
     0.3 & $1.880(1)$ & $-2.76(6)$ &  $-0.916(9)$ & $-0.88(6)$ & $ 2.125(2)$ & $0.816(7)$\\     
    \bottomrule[1.25pt]
  \end{tabular}
\end{center}
\captionof{table}{Numerical value of the different terms in the Shuttleworth equation (see \cref{Def:Shutt}) for (111) orientation with strain applied in the $x$ direction. $\Delta \mathfrak{s}$ represents the surface excess entropy calculated from \cref{Eq:entropy}. }
        \label{Tab:exx111}
\end{minipage}
The internal energy per unit area converges to $\gamma$ for $T=0$ as expected. Based on \cref{Tab:exx111}, the excess entropy for (111) has a weak dependence on temperature and so the difference between excess internal surface energy and $\gamma$ grows approximately linearly with T.
The errors follow the behaviour expected as they increase with the temperature. We want to highlight here that results presented in \cref{Tab:exx111} and SM, while well equilibrated on their own, are the realisation of a single set of initial conditions and a further reduction of this error can be obtained by averaging over simulation using different initial conditions. However, the relatively small error bars we obtained show that the conclusions are consistent with the theoretical model we derived in the previous sections.

\section{Conclusions}

\noindent In this work we presented a unified framework able to describe two important surface properties: Surface Free Energy and Surface Stress. We reported a non-exhaustive literature review where the term surface tension is still in use and we advise against its further use, in favour of the terms ``Surface Free Energy'' and ``Surface Stress'', which are unambiguously defined for all types of surfaces and interfaces involving both liquids and solids. The reason lies in the fact that the term ``surface tension'' and SFE are well defined only for liquids. In solids, the dual use of this term, either meaning the SFE or the SS, often leads to confusion. 

We presented a theoretical derivation for SS and SFE based on the Guggenheim approach, which we deem more suitable for MD simulations where the spatially dependent profiles of various quantities can be directly accessed. 
We then gave an interpretation of the SS in terms of thermodynamic quantities and from this we established its relation with the SFE. In doing that, we showed that the Shuttleworth equations is just the first order approximation of the bespoken relation, which in our opinion settles all the discussion about its validity.
We applied this framework to a face-centered cubic Lennard-Jones crystal at three different orientations, (100), (110) and (100) and four different temperatures, T=0, 0.1, 0.2, 0.3$\epsilon$. In the case of $T=0$ we analytically derived the equations for the SS and its relation with the SFE. The calculations at zero temperature show, as expected, perfect agreement within the numerical error. 
The results at finite temperature come with an associated statistical error but we were able to show the agreements of the results with the theory for every case considered.
The theory just presented was derived under broad hypothesis. However, to make it general, two further generalisations are needed: the case of non-hydrostatically stressed solid and multiple components in the two phases.   These will be the focus of future work.

\section{Acknowledgment}

\noindent This work was part of the  EPSRC grant Virtual Formulation Lab (EP/N025261/1). Authors would like to acknowledge our academic partners: Imperial College, University of Leeds, University of Leicester and University of Greenwich. The simulations are performed at the Tier-2 High Performance Computing centre, HPC Midland Plus, which is gratefully acknowledged.
 
%
\bibliographystyle{plainnat}

\bibliography{bibliography_20Feb20}

\end{document}